\documentclass[journal]{IEEEtran}
\usepackage{hyperref}
\usepackage{datetime}
\usepackage{graphicx}
\usepackage{cite}
\usepackage[T1]{fontenc}
\usepackage[medium]{FiraMono} % Loads the Fira Mono font
\usepackage{listings}
\usepackage[dvipsnames]{xcolor}
\usepackage{float} % Required for [H]

\usepackage{tabularray}
\UseTblrLibrary{booktabs} % Allows \toprule, \midrule, \bottomrule inside tblr
\usepackage{listings}
\usepackage{hyperref}
\usepackage{enumitem} 
\usepackage{siunitx}
\usepackage{comment}
\usepackage{subcaption}
\usepackage{microtype}
\usepackage{tcolorbox}
\usepackage{helvet}
\usepackage{tikz}
\usetikzlibrary{shapes.geometric, positioning,fit,arrows.meta,calc,backgrounds}

\setenumerate[1]{ref=\thesection.\arabic*}

\definecolor{codegreen}{rgb}{0,0.6,0}
\definecolor{codegray}{rgb}{0.5,0.5,0.5}
\definecolor{codepurple}{rgb}{0.58,0,0.82}
\definecolor{backcolour}{rgb}{0.95,0.95,0.92}
\definecolor{mlirtype}{HTML}{4F5B66}

\newfloat{listing}{!tbp}{lol}
\floatname{listing}{Listing}

\usepackage{pifont}
\newcommand{\step}[1]{\smash{\raisebox{-0.35ex}{\fontsize{12}{12}\selectfont\ding{\numexpr171+#1\relax}}}}
\lstdefinestyle{mystyle}{
    backgroundcolor=\color{backcolour},   
    commentstyle=\color{codegreen},
    keywordstyle=\color{magenta},
    numberstyle=\tiny\color{codegray},
    stringstyle=\color{codepurple},
    basicstyle=\ttfamily\scriptsize\bfseries,
    breakatwhitespace=false,         
    breaklines=true,                 
    captionpos=b,                    
    keepspaces=true,                 
    numbers=left,                    
    numbersep=5pt,                  
    showspaces=false,                
    showstringspaces=false,
    showtabs=false,                  
    tabsize=2,
    frame=single,
    xleftmargin=0.05\columnwidth,
    framesep=0.25em,
}

\lstdefinelanguage{MLIR}{
    sensitive=true,
    alsoletter={._},
    morecomment=[l]{//},
    morestring=[b]",
    morekeywords=[1]{
        transport.create,
        transport.connect,
        transport.exchange_keys,
        transport.establish_channel,
        transport.set_message_sizes,
        transport.start,
        transport.get_session,
        transport.stage_payload,
        transport.post,
        transport.reply_slot,
        transport.collect
    },
    morekeywords=[2]{
        transport.session,
        memref,
        index,
        i1,
        i32,
        i64,
        ui16
    }
}

\lstdefinestyle{mlirstyle}{
    style=mystyle,
    basicstyle=\ttfamily\footnotesize,
    keywordstyle=[2]\color{mlirtype},
    numbers=none,
    columns=fullflexible,
    xleftmargin=0pt,
}

\definecolor{lstColor}{HTML}{127882}
\newcommand{\lstIL}[1]{\mbox{\lstinline[columns=fixed, basicstyle=\color{lstColor}\ttfamily\small\bfseries]{#1}}}

\usepackage{amsmath}
\usepackage{url}
\hypersetup{colorlinks=true, allcolors=blue}

\newdate{articleDate}{8}{9}{2026}

\title{Python in the front, party in the Backline: {compiling quantum workloads across CPUs, GPUs, and FPGAs}}

\author{Joseph~K.~L.~Lee,
         Mehrdad~Malekmohammadi,
         Hong-Sheng~Zheng,
         Shuli~Shu,
         Cheick~Doumbia,
         Kalman~Szenes,
         Mehran~Zamani~Abnili,
         Thomas~Ainsworth,
         Matthew~Seymour,
         Thomas~Germain,
         Leonhard~Neuhaus,
         Josh~Izaac,
         and~Lee~J.~O'Riordan
 \thanks{All authors are affiliated with Xanadu Quantum Technologies Inc.,
 Toronto, Ontario M5G 2C8, Canada (e-mail: backline@xanadu.ai).}}

\begin{document}
\maketitle

\IEEEtitleabstractindextext{%
\begin{abstract}Moving from quantum research and development to production-grade, fault-tolerant quantum workload execution remains one of the most significant challenges facing quantum platform builders. 
While Python frameworks have enabled an easy entry point for quantum algorithm design, the low-latency requirements for real-time quantum error correction (QEC) demand performance that traditional interpreted environments cannot provide. FPGAs and ASICs play a central role at these layers, but their specialized programming models make development rigid and time-consuming. CPUs, GPUs, and other accelerators introduce a different challenge: as infrastructure becomes increasingly heterogeneous, programming across different devices and their associated abstractions becomes more complex.
Designing abstractions that allow researchers to write workloads in high-level languages that map to low-latency execution across diverse distributed target platforms will enable the development of key infrastructure for utility-scale quantum systems. For this, we introduce \textit{Backline}, a heterogeneous compilation and runtime framework built within PennyLane and Catalyst. {Backline allows us to design and build quantum-classical workloads for high-performance and low-latency devices, with compilation directly from a Python interface through MLIR.} We demonstrate the compilation and execution of several quantum workloads with low-latency data movement across a mix of CPUs, GPUs, and FPGAs, efficiently handled for both local and distributed remote hardware targets, all from a vendor-agnostic Python frontend.
With an AMD VPK120 FPGA board as the controller, issuing each round from its hardware-handshake engine, we measured median steady-state round-trip latencies over RoCE~v2 of \qty{2.305}{\us} to an AMD Ryzen Threadripper PRO CPU and \qty{4.5}{\us} to an AMD Instinct MI210 GPU across $10^6-1$ rounds per path, demonstrating microsecond-scale synchronous co-processing.
\end{abstract}
}

\maketitle
\IEEEdisplaynontitleabstractindextext

\section{Introduction}

Across many domains of scientific computing and research software, writing performant and expressive systems is a complex endeavour.
Allowing domain experts to work at the right layer of abstraction enables them to efficiently design, prototype, and build out workloads that move their research forward. A quantum computing algorithm researcher should not need to understand memory access optimizations to implement their algorithms, as such low-level details can and should be abstracted away. Similarly, an HPC performance engineer should not need to understand the complex mathematical formalism of the special unitary group, only that the data is handled efficiently on the underlying hardware.

In the field of quantum system design, we have been able to avoid such targeted complexities in the past with noisy intermediate-scale quantum (NISQ) machines, since fast feed-forward from measurement outcomes to subsequent corrections was never a part of the execution paradigm. For fault-tolerant quantum hardware environments we now need to take this into consideration, as the application of gates based on branched pathways is a fundamental part of the execution paradigm. Quantum error correction (QEC) for fault-tolerant quantum computing (FTQC) has extremely tight communication and feedback loops.  Additionally, as we require quantum devices to be programmable and controllable in such low-latency environments, FPGAs and ASICs become paramount components of the stack. Writing software for devices at these layers remains hard and rigid, as ASICs are often built to serve a specific singular purpose, and resynthesizing FPGA bitstreams can be a time-consuming process for large workloads~\cite{fpgaRoute2013}. As QEC requires researchers and quantum control-system developers to interact with these devices, appropriate software abstractions can help hide unnecessary implementation complexity. At the same time, researchers designing algorithms and applications often have heavy computational requirements, which on the spectrum of quantum workloads can be compute-bound, communication-bound, and everything in between~\cite{rao2026performancemodelhybridquantumclassical}.

Since FTQC systems remain under active development across all quantum hardware platforms, there remains significant R\&D for software, hardware, QEC, and architecture, with many identifying various requirements across the stack for both applications and infrastructure~\cite{shehata2026quantumhpcsoftwarestacksopenqse}. 
For fast progress, we need stacks that support this R\&D and co-design end-to-end with layered interfaces for everyone involved, from application developers to system engineers. This means supporting users who are prototyping FTQC workloads with QEC in high-level languages (e.g. Python), and allowing them to progressively compile and optimize these workloads onto local and distributed heterogeneous devices matching the real-world environments and components in which large-scale FTQC systems will operate.

As these systems scale, developing software that spans both distributed computing and quantum computing becomes increasingly complex, and this complexity is a recognized challenge~\cite{mohseni2026buildquantumsupercomputerscaling, iris2024, fur2026opportunitieschallengesscalingquantum, seelam2026referencearchitecturequantumcentricsupercomputer, shehata2026quantumhpcsoftwarestacksopenqse}. 
Within fast-moving technical fields, two disparate approaches are often taken: i) aim to lock down and standardize the stack to allow multiple parties to unify around a common base, or ii) move fast with less-than-optimal designs to unblock important ecosystem research work~\cite{mlTechDebt2025}. While both have their merits, stability and performance can often be at odds, requiring a nuanced approach in research and design, and standardizing a fast-moving field too early can entrench suboptimal solutions~\cite{riseFallCorba2008, osiVsTcp2013}. Platforms for QEC infrastructure (both hardware and software) remain highly bespoke, with limited access to interface layers allowing development against multiple hardware variants. While efforts like NVQLink~\cite{nvqlink2025} are a move in the right direction, the lack of variety in target platforms is a recognized constraint. Indeed, having multiple vendor-agnostic platforms would be valuable to the community~\cite{chen2026realtimequantumerrorcorrection}. Therefore, we favour a moving-design specification that evolves with the field, rather than a rigid standard, unlocking the best of both approaches above. We approach these systems with ``best effort'' compatibility, similar to the LLVM ecosystem's approach to stability and compatibility between releases~\cite{llvmDevPolicy2026}, letting us move quickly with researcher and ecosystem needs while still offering compatibility guarantees across changes.

Recent trends in software development for quantum ecosystems follow the early days of building expansive machine-learning pipelines, with increasing development, abstraction, and sophistication across the stack. As workload complexity expands, the need for managed device interactions and distributed environments is becoming central to the ecosystem's design. 
The need for communication across multiple hardware and system-level components, and doing so with minimal overheads, is at the core of many such ecosystems. For example, consider machine learning inference workloads, where disaggregated computing environments assign dedicated devices to dedicated functions and behaviours~\cite{hpcDisagg2024, disaggDataCenterArch2026}. In this scenario, a variety of device types with unique responsibilities are treated as resources, and are engaged according to behaviour that matches a given workload's needs. 

Frameworks like TensorFlow, PyTorch, and JAX~\cite{tensorflow2016, pytorch2019, jax2018} have enabled rapid progress in abstracting away much of the related machinery and configuration needed to scale up and scale out ML workloads. In each of these, Python is the frontend, with support for both eager (Python interpreter) and compiled workloads, allowing execution on local and remote CPUs, GPUs, TPUs, and custom devices. As these frameworks follow a familiar NumPy-API design for writing user algorithms, applications, and end-to-end workloads, they have also expanded the infrastructure to support extensibility and customization across the ecosystem. Kernel-development tooling like Pallas~\cite{pallasJax} and Triton~\cite{triton2019} expand the above SDKs to allow explicit definition of complex accelerator kernels, directly from Python. The usability and performance of these tools, relative to bare C-style extensions, have proven to be advantageous for compute-heavy workloads as well as for productivity~\cite{tritonVcuda2026}. As workloads expressed in such frameworks scale, additional consideration must be given to data locality and availability for all devices in the system. Many frameworks utilize an explicit data-placement model, allowing data to be manually migrated between devices as dictated by the workflow needs. To reduce bottlenecks in such workloads, this migration needs to happen as efficiently as possible.
Key technologies to support this are the commonly used direct memory access (DMA) abstractions, where direct local communication between devices can happen on a given node, and remote memory access via RDMA~\cite{roce2016, hpcRDMA2016} where communication between devices across different nodes is enabled by efficient device-to-NIC communication. While programming with low-level abstractions is common in HPC ecosystem design, application developers should not need to understand, or even be aware of, these complex system layers. Given that quantum workloads will largely run on mixtures of commodity, HPC, and bespoke hardware across various component types, it is necessary to create well-designed abstractions for the stack across all layers.

For programming quantum devices, the use of high-level Python programming APIs backed by MLIR/LLVM compiler infrastructure is becoming the de facto design of choice to abstract away the complexity of the executing environment from the user-facing layers~\cite{pennylane2018, Ittah2024, ibmMLIR2024, mlirCatMQT2026, cudaq2025}.
In PennyLane, a program is written in Python, and compiled by the Catalyst compiler through an MLIR and LLVM pipeline to progressively lower the classical and quantum program to device-native operations. While MLIR and LLVM are expansive compilation tools, they complement the Python-first designs, helping to keep the complex system and compilation layers abstracted away from quantum application developers. 
This compilation-first approach unlocks many abilities that have limited exposure in NISQ-first stacks, such as preservation of program structure~\cite{ittah2025constanttimehybridcompilationshors}, complex targeted decompositions, and hierarchical optimization, as well as integration of existing classical code in the same program.

Given the separation between high-level algorithmic research for optimal program representations, and high-performance low-latency workload execution, marrying both across a variety of device types can be a daunting challenge. Our manuscript will address this exact scenario. We will explore how to provide accessible high-level programmability while maintaining best-in-class low-latency performance for quantum workloads through a heterogeneous compilation and runtime layer we call \textit{Backline}.

The manuscript is structured as follows: 

\begin{itemize}
    \item Section~\ref{section:qec} provides a snapshot of fault-tolerant quantum computing, hardware constraints, and R\&D requirements.
    \item Section~\ref{sec:programming} describes the open-source software PennyLane for quantum programming and Catalyst for compiling the resulting programs to heterogeneous hardware devices.
    \item Section~\ref{sec:lowlatency} focuses on the design of Backline's interoperability between heterogeneous devices and low-latency communication abstractions from PennyLane through Catalyst.
    \item Section~\ref{sec:workloads} demonstrates usage of Backline abstractions via PennyLane and Catalyst through examples of low-latency workloads running across asymmetric heterogeneous devices.
    \item Section~\ref{sec:conclusions} examines the applicability of the infrastructure for both current and future FTQC workloads.
\end{itemize}

\section{FTQC platforms and QEC}\label{section:qec}

In the current state of quantum computing infrastructure, one active research topic of high importance is the development of QEC protocols for encoding, workload execution, and decoding of errors \textit{in real-time}.

For Xanadu's Aurora system~\cite{Aurora2025}, the error correction decoder computation was implemented on an FPGA with a latency of \qty{64}{\ns}. The end-to-end feed-forward latency within which the decoder needed to provide corrections back to the system was slightly below the available \qty{1000}{\ns} budget due to additional latencies coming from analog, analog-to-digital, and serial communication delays imposed by the hardware selected for the experiment. More optimal hardware choices such as using zero-latency analog-to-digital converters and optimized serial transceivers can bring this additional latency below \qty{100}{\ns}. As this example relied on a small-scale code, increasing the complexity of the QEC schema will further extend the execution times of the decoder, potentially exceeding the latency budgets for real-time corrections. There is currently significant ongoing work to explore hardware implementations that enable faster decoding in small-time windows~\cite{parallelBPOSD2026, scalablefpgaarchitecturerealtime2026}.

Considering the decoding loop of the Aurora system, where an inner and outer decoder are paired to ensure logical correctness of the system within a given number of clock cycles, having the ability to prototype such protocols can help unlock an accessible R\&D path for system-level research. Because these quantum hardware ecosystems are highly complex, building infrastructure that supports the iterative design and testing of these protocols is essential. This infrastructure enables teams to understand bottlenecks, application viability, and overall system-level requirements for real-world quantum workloads. Consequently, a unified end-to-end stack (programming, compilation, orchestration, runtime execution) becomes paramount for effective hardware-software co-design, ensuring that classical control architectures and quantum processors are iteratively tailored to mutually support one another's constraints.

However, this co-design process must account for highly non-uniform system requirements. The various classical tasks surrounding a quantum device operate under vastly different timing deadlines, and conflating them obscures which specific classical hardware is actually needed at each layer. For the purpose of quantum system hardware, we can loosely categorize latencies into different regions of utility. It is important to separate two distinct axes: i) what a given class of classical hardware can deliver, and ii) what a given quantum modality actually requires. We first lay out the capabilities and what is achievable, and then map the per-modality requirements onto it.

Following the resource analyses of real-time decoding and hybrid
quantum-classical execution~\cite{Battistel_2023, rao2026performancemodelhybridquantumclassical},
we define the following tiers of latency-bounded operation for an FTQC system.
These tiers describe the class of hardware that can service a given latency
band, with the modality then dictating which tier must be met, and looser tiers enabling more complex computations at lower cost and development effort:
\begin{enumerate}[ref=Tier~\arabic*]
\item \textit{Real-time control tier}\label{enum:realtime} [\qty{1}{\ns}--\qty{1}{\us}]: The tightest latency regime, servicing hard real-time tasks such as QEC decoders and associated adaptive control hardware, where each computation must complete within a deterministic, worst-case-bounded budget~\cite{Battistel_2023}. ASICs, FPGAs, and embedded real-time coprocessor cores are the primary targets at this scale. 
\item \textit{Synchronous co-processing tier}\label{enum:synchronous} [\qty{1}{\us}--\qty{100}{\us}]: A soft real-time regime for supportive, high-importance tasks such as backup decoding and auxiliary processing that can tolerate some variation in execution time. Coherent memory access and synchronous low-latency kernels on FPGAs, CPUs, and GPUs can serve this regime, depending on the computational requirements.
\item \textit{Co-located runtime tier}\label{enum:runtime} [\qty{100}{\us}--\qty{1}{\s}]: For batched or numerical assistive workloads, as well as tight compilation loops. Execution here should not sit on the critical path for quantum workloads, and can draw on externally available libraries and tooling, such as high-throughput communication collectives and optimized linear-algebra implementations.
\item \textit{Distributed workload tier}\label{enum:distributed} [$>$\qty{1}{\s}]: Standard data-processing workloads, where start-up costs can be paid to allow subsequent communication and compute to overlap effectively. This covers pre- and post-processing, as well as non-critical calculations on local or remote systems.
\end{enumerate}

The tier that a system is required to hit is set by the QEC cycle rate of the underlying qubit modality, since the decoder throughput must keep pace with the incoming syndrome data; otherwise the computation suffers an exponential backlog slowdown~\cite{Battistel_2023}. The required timescales span several
orders of magnitude across platforms, where tighter modalities such as photonic and superconducting systems sit around \qty{1}{\us}, while looser ones such as ion traps and neutral atoms sit closer to \qty{1}{\ms}. For photonic systems, \ref{enum:realtime} is a hard requirement for decoding. 
That budget binds the primary decoder while the looser tiers exist for the classical work that does not need to keep pace with the cycle rate.

For workloads that have a hard deadline on the arrival of a result, strategies such as latency hiding do not provide any benefit: if the result is
unavailable, the outcome can be fatal to the workload. Running programs on real-world quantum hardware falls into this category, with
error-correction subroutines being essential to ensure the functional operation of the device.
Moreover, iterative decoders such as belief propagation (BP), are inherently probabilistic, which means that there is a given likelihood that the decoder does not converge in the operation timescale budget.
This leads to a stall of the system until the correction can be determined and applied.

While an FPGA decoder is expected to correct the majority of such errors, some small tail of events (e.g. those beyond the {P99.999} percentile) cannot be corrected in a timely manner and introduce the problems mentioned above. These tail-errors are too complex to converge with a fixed decoder design, and may, therefore, benefit from an additional auxiliary processor equipped with a more sophisticated algorithm. Such a decoder is not required on every round, and, therefore, does not need to meet the real-time control tier budget above. Indeed, it belongs in \ref{enum:synchronous}, which is reachable with commodity hardware. These decoders could run on a variety of heterogeneous devices (CPU, GPU, or even another FPGA), and assist the system to handle low-likelihood tail-errors~\cite{decoderSwitching2025}, preventing the decoding backlog from building up to unresolvable amounts~\cite{RevModPhys.87.307} and allowing execution to continue.

To minimize stalls, this auxiliary hardware should i) have access to the relevant data, ii) be primed for immediate execution of the decoder, and iii) be ready to transmit the correction back for application. Realizing these concepts calls for careful consideration of both the user interface and user experience, as well as the underlying infrastructure and protocols needed to support workloads operating under this paradigm. 
For these abstractions to be practical on real-world quantum systems, they must be supported by a low-latency execution environment, particularly given the communication overhead associated with transferring data. At the same time, the system should retain sufficient programmability to enable a reasonably short turnaround from prototyping to experimental data collection.

\section{Programming and compiling quantum programs}\label{sec:programming}

\begin{figure}[!htbp]
	\includegraphics[width=\columnwidth, clip, trim=4.5cm 16.5cm 25cm 2cm]{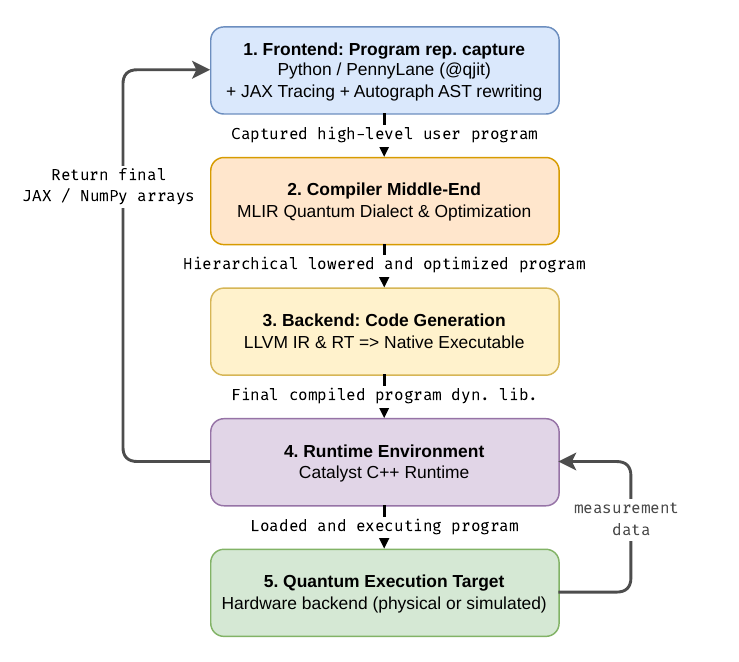}
	\caption[]{High-level overview of the PennyLane program lowering process.}
	\label{fig:pl2cat}
\end{figure}

High-level software abstractions are needed to hide the complexity of the underlying quantum hardware. To accommodate the increasing heterogeneity in computing systems, many frameworks have been developed to provide this functionality at various layers of abstraction, with Pythonic frontends coupled with compiled backends being widely used~\cite{pennylane2018, Ittah2024, javadiabhari2024quantumcomputingqiskit, seidel2024qrispframeworkcompilablehighlevel, Koch2024, cudaq2025}. Allowing for the user program to be written in Python and natively compiled on a variety of hardware types (e.g. CPUs, GPUs, TPUs, QPUs, etc.) helps remove execution overheads from Python interpreters and schedule execution on the most effective device target~\cite{torch2_2024, jax2018}. 
PennyLane's compiler \textit{Catalyst} allows for the expression of workloads at the Python layer, with the program progressing through multiple lowering stages from MLIR, onto LLVM IR, and to a final binary program representation for the executable workload. Quantum-device-specific calls are provided by an associated runtime layer, which is linked against the program for execution on the target hardware.
For a PennyLane program, compilation through Catalyst follows the high-level design given by Fig.~\ref{fig:pl2cat}.

Since programmability is the most important enabler of any new technology (if you cannot program it, you cannot use it), defining the right abstractions becomes important for end-users and researchers. Shallow-embedded DSLs (e.g., tracing interpreters) are often favoured for their extensibility over deep-embeddings (e.g. AST ``interpreters'') due to the host language and associated libraries being implicitly accessible to provide key functionality as expressed by the user~\cite{yinYangDSL2014}. This usability comes with a trade-off: deeper embeddings provide more compiler-friendly constructs by enabling direct interpretation at the AST-level. To ease the extensibility of our ecosystem, we opt for a dual-mode approach comprising a shallow embedded DSL~\cite{edsl2012} that is traced to capture the intention of the quantum program, with additional functionality provided by an AST-level rewriter that captures and represents Python's structured control-flow~\cite{autograph2019}.

In this way, we treat Python as the language to express the user's intent and to build the compiled program, rather than the interpreted execution environment it is commonly used for. Within this environment, users can express their workloads for heterogeneous compilation and execution, with our \lstIL{qjit} frontend layer handling the capturing and lowering of the program from high-level concepts through MLIR all the way down to the final compiled program. The PennyLane tracing mechanism is currently based on JAX~\cite{jax2018, jax2018github} and, thus, relies on similar infrastructure to represent the high-level program, before explicit conversion to MLIR, from which Catalyst~\cite{Ittah2024} lowers the program to the pre-defined targets.

\section{Low-latency computing with HPC-grade hardware}\label{sec:lowlatency}

While Section~\ref{sec:programming} focused on programming abstractions and compilation, one must also consider a runtime layer for such compiled workloads, handling the direct tie-in to the hardware, as well as data migration, synchronization, and related system-level functionality. Operating systems often have well-defined
abstractions that allow many devices to communicate effectively with the operating system kernel, behaving as a mediator in all communication paths. These
layered abstractions allow for an effective way to build interoperability between devices, treating the system-provided abstractions as the
means to ``wire'' devices together in flexible ways.

Such flexibility however can come at the cost of performance: ensuring that there is always a mediator handling the data as it traverses the system may
add some overheads which can be reduced by a more direct communication path. To meet the real-time needs of quantum hardware systems, it becomes important to design software ecosystems with data placement and mobility built-in across all abstraction layers for low-latency execution.

Of the tiers set out in Section~\ref{section:qec}, \ref{enum:realtime} is reachable only with bespoke hardware infrastructure sitting beside the quantum device. We therefore focus on programmability within \ref{enum:synchronous}, with \ref{enum:runtime} and \ref{enum:distributed} also implicitly supported.

For \ref{enum:synchronous}, having easily programmable auxiliary devices to complement the real-time control path allows for more flexible development, design, and exploration of key infrastructure, such as auxiliary decoding and switching~\cite{decoderSwitching2025}. Building such low-latency infrastructure, as well as the workloads that run on them, relies on many approaches to reduce overheads, with kernel bypass techniques being commonly used.
These techniques allow user-space applications to skip over standard processing practices that happen in the Linux kernel,
allowing the removal of overheads by more direct data communication between hardware. We can consider the use of remote memory access paradigms, allowing data exchange between classical devices in the most effective manner based on the topology and composition of the target system. The use of such paradigms has recently been both discussed~\cite{rdmaqec2025} and demonstrated~\cite{nvqlink2025} for quantum programming environments, tying into the remote direct-memory access (RDMA) subsystem of the Linux operating system and using {RDMA over Converged Ethernet} (RoCE)~\cite{roce2016} as the underlying network and transport protocol.

Through this, costly context switches from the CPU can be avoided, as well as unnecessary memory movement between PCIe-accessible devices and main memory.
Instead, PCIe devices communicate their messages directly while minimizing system-level involvement.
Assuming a user wishes to read from or write to a remote target's memory, the accelerators and the RNICs are the only participants in the data exchange, avoiding costly CPU and DRAM accesses that are normally incurred in default execution pathways.

This infrastructure is used by most, if not all, HPC systems, as well as for running large-scale machine-learning workloads~\cite{rdmaInAI2024}, where both high throughput and low-latency data transmission are essential for reducing time-to-solution. 
In large-scale quantum systems, using these networks for auxiliary decoding requires advanced scheduling policies to allocate decoder resources efficiently~\cite{elasticDecoder2026}. For most R\&D applications, however, a tightly connected mesh of devices is sufficient. Supporting these use cases requires mechanisms for expressing communication paths between devices with different target architectures, as well as runtime layers that abstract data migration from end users.

\subsection{Backline}
With the above needs identified, we introduce \textit{Backline}, a high-level data-placement abstraction, heterogeneous compilation layer, and low-latency runtime layer, built into PennyLane and Catalyst.

\begin{figure}[!htbp]
\centering
\def\bandw{0.74\columnwidth}   % content width shared by all three bands
\def\subw{0.68\columnwidth}    % indented content inside a band
\def\bandc{0.37\columnwidth}   % centre of a band == \bandw/2
\def\boxw{0.30\columnwidth}    % ONE width for all six inner boxes
\def\ind{3.2mm}                % indent of a band's contents from its header
\def\gap{8mm}                  % vertical gap between bands (sets arrow length)
\begin{tikzpicture}[
  font=\sffamily\scriptsize,
  role/.style={draw=black!45, rounded corners=1pt, fill=white, align=left,
               text width=\boxw, inner sep=3pt},
  area/.style={rounded corners=3pt, inner sep=3.5pt},
  hdr/.style={align=left, text width=\bandw, anchor=north west},
  sub/.style={align=left, text width=\subw, anchor=north west},
  note/.style={black!55},
  flow/.style={-{Stealth[length=2.2mm]}, thick, black!55},
  wire/.style={{Stealth[length=1.8mm]}-{Stealth[length=1.8mm]}, thick, Red!70},
]
\coordinate (O) at (0,0);          % band left edge
\coordinate (CX) at (\bandc,0);    % band centre, for centring box pairs
\coordinate (IX) at (\ind,0);      % band indent

\node[hdr] (plchdr) at (0,0)
  {\textcolor{blue!75}{\textbf{Placement}} \;--\; \texttt{qp.Backline(...)}%
   \hfill \textcolor{blue!75}{\textbf{PennyLane frontend}}};

\node[sub] (nodehdr) at ([xshift=\ind, yshift=-1.4mm]plchdr.south west)
  {\textcolor{blue!65}{\textbf{Node}} \;--\; target, locality, connectivity};

\node[role, anchor=north east] (ctrl) at ([xshift=-1.5mm, yshift=-1.4mm]nodehdr.south -| CX) {%
  \textbf{Controller}\\[1pt]
  drives the QNode,\\initiates every round\\[1pt]
  \textcolor{black!60}{CPU \textbar{} FPGA}};
\node[role, anchor=north west] (coproc) at ([xshift=1.5mm, yshift=-1.4mm]nodehdr.south -| CX) {%
  \textbf{Coprocessor}\\[1pt]
  runs a coprocessor\\function per message\\[1pt]
  \textcolor{black!60}{CPU \textbar{} GPU}};

\node[sub] (transport) at ([yshift=-3.6mm]ctrl.south -| IX)
  {\textcolor{blue!65}{\textbf{transport}} \;--\; \texttt{"memcpy"} or \texttt{"rdma"}};

\begin{scope}[on background layer]
  \node[area, fill=blue!9, draw=blue!45, fit=(plchdr)(ctrl)(coproc)(transport)] (plcbox) {};
  \node[area, fill=blue!3, draw=blue!35, fit=(nodehdr)(ctrl)(coproc)]           (nodebox) {};
\end{scope}

\node[hdr] (cathdr) at ([yshift=-\gap]plcbox.south -| O)
  {\textcolor{OliveGreen!65!black}{\textbf{MLIR dialects}}%
   \hfill \textcolor{OliveGreen!65!black}{\textbf{Catalyst compiler}}};

\node[role, anchor=north east] (exec) at ([xshift=-1.5mm, yshift=-1.4mm]cathdr.south -| CX) {%
  \textbf{executor}\\[1pt]
  which device instructions\\run on, and how it\\gets there};
\node[role, anchor=north west] (tport) at ([xshift=1.5mm, yshift=-1.4mm]cathdr.south -| CX) {%
  \textbf{transport}\\[1pt]
  how data moves\\between them, once\\they are running};

\begin{scope}[on background layer]
  \node[area, fill=OliveGreen!8, draw=OliveGreen!45, fit=(cathdr)(exec)(tport)] (catbox) {};
\end{scope}

\node[hdr] (rthdr) at ([yshift=-\gap]catbox.south -| O)
  {\textcolor{orange!75!black}{\textbf{unified C ABI}}%
   \hfill \textcolor{orange!75!black}{\textbf{runtime}}};

\node[role, anchor=north east] (rtc) at ([xshift=-1.5mm, yshift=-1.4mm]rthdr.south -| CX) {%
  \textbf{controller target}\\[1pt]
  session life cycle,\\memory regions,\\drives each round};
\node[role, anchor=north west] (rtd) at ([xshift=1.5mm, yshift=-1.4mm]rthdr.south -| CX) {%
  \textbf{coprocessor target}\\[1pt]
  binds and runs the\\coprocessor function\\for each message};

\draw[wire, rounded corners=1.5mm]
  (rtc.south) -- ([yshift=-4mm]rtc.south) -| (rtd.south);
\node[note, anchor=north] (wlbl)
  at ([yshift=-4.6mm]$(rtc.south)!0.5!(rtc.south -| rtd)$) {rounds};

\begin{scope}[on background layer]
  \node[area, fill=orange!10, draw=orange!55, fit=(rthdr)(rtc)(rtd)(wlbl)] (rtbox) {};
\end{scope}

\draw[flow] ([xshift=9mm]plcbox.south west) -- ([xshift=9mm]catbox.north west)
  node[note, midway, right=1.5mm] {carried into the compiler};
\draw[flow] ([xshift=9mm]catbox.south west) -- ([xshift=9mm]rtbox.north west)
  node[note, midway, right=1.5mm] {lowered to the runtime's C entry points};

\end{tikzpicture}
\caption{The Backline abstractions and how they compose.}
\label{fig:backline_overview}
\end{figure}
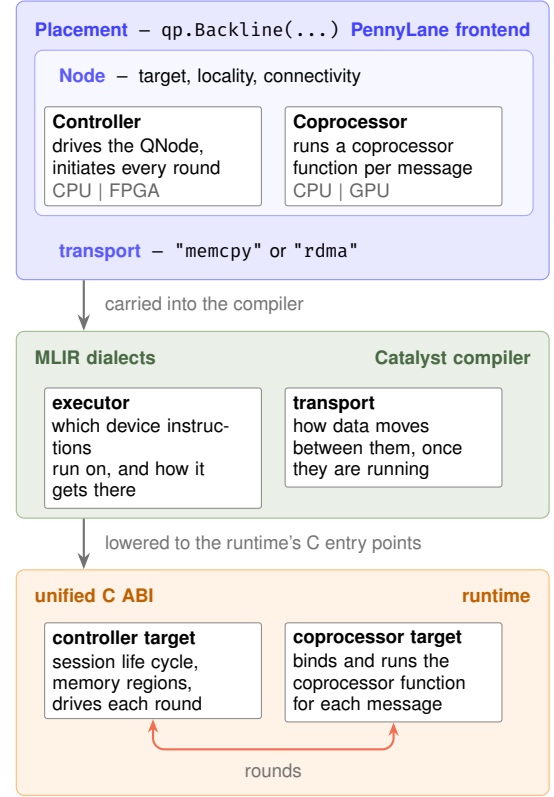

Backline spans three layers: 1) The \textbf{PennyLane frontend} provides a declarative interface for specifying the target devices that participate in a workload, assigning computations to those devices, and selecting the transport protocol used for data exchange. 2) The \textbf{Catalyst compiler} translates this specification into MLIR, compiles program components for their assigned execution environments, and coordinates their deployment and execution. 3) The \textbf{runtime} implements target-specific execution, memory management, and data movement, including registering memory regions, binding the function that a coprocessor runs, and actually moving payloads between the devices. This modular setup enables easy development of new runtime devices with a unified and portable user interface. Figure~\ref{fig:backline_overview} shows how the various components compose.

\subsection{User interface for workload expression}\label{sec:backline_ui}
Our aim for the frontend layer is to enable users to prototype a heterogeneous workload locally on a laptop with a CPU, or on a single workstation with a GPU, which can subsequently be moved onto HPC-grade hardware with an RDMA-capable NIC to achieve very low latency with minimal changes to the program, all directly within Python through PennyLane.

\subsubsection{Data placement abstraction}
Data movement is primarily implicit in the model. The user creates a heterogeneous device explicitly through the usual PennyLane interface, and the transfers between the component targets follow from the declaration, rather than being written out by hand. The abstraction at this layer comprises the following components:

\begin{itemize}
    \item A \lstIL{Node} is one participant in a (heterogeneous) workload.  It carries the information needed to identify and reach that participant: the \lstIL{hardware} it executes on, which the compiler pairs with the placement's transport to resolve a concrete target library; whether it is \lstIL{remote}; how its code is deployed there; and where the role requires, its connectivity and arguments to set it up. The following two roles specialize this:
    \item A \lstIL{Controller} is the concrete root of the placement model and drives the quantum device: it initiates all workload execution, mediates data transfers, and is directly responsible for returning results to the originating user process once the compiled workload completes. This could be a CPU, or an FPGA, and it executes a \lstIL{QNode}.
    \item A \lstIL{Coprocessor} is an accelerator that assists the controller, for example as a decoder. This can be a CPU, a GPU, or any platform directly targetable with the supported compiler and backend infrastructure. Each one registers a \textit{coprocessor function}, which defines what runs there for every message it receives.
    \item A node may run in the host process, in a local executor process, or in an executor on another machine. Executor-dispatched nodes are compiled for the executor’s target and executed through LLVM ORC, as described in Section~\ref{sec:executor_dialect}.
    \item One controller, zero or more coprocessors, and a transport mode together define a \lstIL{Placement}, declared with \lstIL{qp.Backline}. This is the control plane, i.e. the single place a workload's topology is defined. \lstIL{qp.Backline} creates a heterogeneous device, which binds directly to a PennyLane QNode and carries the placement into the compiler.
\end{itemize}

Taken together, this is an implicit memory model for interactions between target devices. Explicit memory interactions are commonplace in ML workloads (e.g. Torch and JAX expose tensor placement directly); here we aim instead for a data model managed by the device mesh itself, so that transfers follow from how its nodes are declared to interact. Concretely, a placement is a single \lstIL{qp.Backline} call carrying a \lstIL{qp.Controller}, a list of \lstIL{qp.Coprocessor} nodes, and a transport mode, with each node holding the fields above. Section~\ref{sec:v_placements} gives complete placements for each of the demonstrated configurations.

\subsubsection{Coprocessor functions}\label{sec:coproc_fn}
A coprocessor function is what a coprocessor applies to each message it receives. It can be defined in several ways, each trading simplicity against control and performance. For the most optimized implementations, the function can be written and pre-compiled directly in a library (e.g. in C++), for either a CPU or GPU target, and referenced by symbol. This gives direct control over the decoder’s inner loop and memory-access patterns.

Alternatively, the function can be written in Python; the provided helper functions take either custom Triton functions, or Tanner graphs for CSS codes, and compile ahead-of-time as GPU kernels that the runtime target can load and launch. This significantly lowers the barrier to prototyping performant GPU decoders directly from Python. Listing~\ref{lst:coproc_fn} shows all three routes.

Additionally, a coprocessor may register multiple functions, dynamically dispatched depending on the incoming message. A section of the payload is allocated for a \lstIL{decoder_id}, which is used to select which decoder function (e.g. for correcting X-type or Z-type errors) is applied to the incoming syndrome.

\begin{listing}[H]
\begin{lstlisting}[language=Python]
import pennylane as qp
import triton.language as tl

# (a) Pre-compiled: name a symbol in a library the
#     runtime can reach. Demos 1, 1a on a CPU;
#     demos 3, 4 on a GPU.
steane_decode = qp.CoprocessorFunction(
    name="steane_coprocessor",
    lib_path=STEANE_CPU_DECODER_LIB_PATH)

# (b) From a Triton kernel, written as plain Python
#     each entry is compiled by the helper. 
#     Demo 5, abridged here.
def steane_lookup(syndrome):
    idx = tl.cast(0, tl.uint32)
    for i in tl.static_range(3):
        idx |= tl.cast((syndrome >> (8 * i)) & 1,
                       tl.uint32) << i
    qubit = (tl.cast(STEANE_LUT, tl.uint32)
             >> (idx * 4)) & 0xF
    return tl.where(qubit == 0xF, NO_ERROR,
                    tl.cast(qubit, tl.uint64))

# decoder_id indexes the tuple at run time, so one
# coprocessor serves X and Z. Steane is self-dual,
# so the one kernel is given twice.
steane_triton = qp.backline.triton_decoder(
    (steane_lookup, steane_lookup),
    platform="hip:gfx90a:64")

# (c) From a Tanner graph: supply the CSS parity-
#     check matrices only. Demos 2, 2a use the
#     [[13,1,3]] hypergraph-product code, whose
#     Hx and Hz are built in the demonstration.
bp_osd_decoder = qp.backline.css_bp_decoder(
    Hx, Hz, postprocess="osd", num_iters=10,
    platform="hip:gfx90a:64")

coprocessor = qp.Coprocessor(
    hardware="gpu", remote=True,
    coprocessor_fn=bp_osd_decoder)
# or coprocessor_fn=steane_decode/steane_triton
\end{lstlisting}
\caption{The three routes to a coprocessor function: naming a pre-compiled symbol, compiling a
user-supplied Triton kernel, or generating a belief-propagation decoder with ordered-statistics post-processing (BP-OSD) decoder from a CSS code's Tanner graph. All
three produce a \lstIL{CoprocessorFunction} and are interchangeable at the placement level. The
comments name the demonstrations of Section~\ref{sec:v_demos} that take each route.}
\label{lst:coproc_fn}
\end{listing}

\subsubsection{Transport mode and local prototyping}\label{sec:transport_mode}
Prototyping locally requires nothing beyond a Linux machine, and optionally a GPU. By selecting a local transport (\lstIL{memcpy}), the heterogeneous compilation and execution path, including placement, serialization, and the per-round request/reply structure, is still exercised. The messages are transferred as local memory copies. When a user has access to a more advanced RDMA-ready setup, changing the transport mode to \lstIL{rdma} replaces those copies with one-sided network writes. No change to the program is required besides updating the transport name and nodes' locality. 

\subsubsection{Granular control and runtime call}
The abstractions above are implicit: the user declares a placement, and the compiler decides where a round of communication appears. For iterative R\&D, this is sometimes more abstraction than one wants, so we also introduce functionality that exposes the runtime directly through \lstIL{qp.runtime_call}, which embeds into the lowered program, at the point of expression, a call to a pre-compiled and linked dynamic library.

This provides the simplest accessible abstraction for calling into performance-optimized programs directly from PennyLane. The dynamic library is loaded by the runtime executor and launched to support the given workload on the target device, which may be either local or remote. It can be used, for example, to transfer single messages between devices, or to trigger a handshake, initialization, or setup steps for a new device. Everything from BLAS libraries, to custom compiled functions can be loaded and executed here, with the data available at the instruction point during workload execution.

\subsection{Compiler infrastructure}\label{sec:hetero_compile}
To ensure that a program can run on multiple hardware platforms, it is often necessary to utilize platform-agnostic programming abstractions and multiple separate compiler toolchains, each targeting a given platform. The LLVM ecosystem as a whole supports compilation for a wide variety of hardware (CPU, GPU, and custom devices), and so is naturally positioned as the compiler infrastructure base for many vendor ecosystems (e.g. AMD's ROCm compiler hipcc, Intel's oneAPI DPC++, etc.). As Catalyst is built atop MLIR, and hence LLVM, we can tie into this tooling with minimal effort, allowing a program to be written and compiled on a given system, and executed on another target architecture and instruction set through cross-compilation. 

Through PennyLane's \lstIL{qjit} interface, users can configure target-specific compilation options and use a \lstIL{Placement} definition to assign specific parts of their program to different devices. To translate these high-level instructions into the intermediate representation (IR), we introduce two complementary dialects: the \lstIL{executor} dialect handles the movement of \textit{code} (determining which target device a subprogram runs on and how it gets there), while the \lstIL{transport} dialect handles the movement of \textit{data} (specifying the exchanges between those subprograms during execution).

\subsubsection{The \lstIL{executor} dialect}\label{sec:executor_dialect}
The placement specifies both the execution process and the machine locality of each program
component. A component may execute in the host process, in a separate executor process on the same machine, or in an executor on another machine. During lowering, components assigned to an executor are compiled separately, and the \lstIL{executor} dialect represents the transfer and invocation of the resulting code. The placement also supplies the compilation target and indicates whether supporting libraries are resolved from the local installation or from a deployed software bundle.

This remote execution life cycle follows a sequence of dialect operations. First, a session is opened against an executor endpoint (\lstIL{executor.open}). Next, a compiled object is shipped to it (\lstIL{executor.send_binary}), and entry points within that object are invoked (\lstIL{executor.launch}) using operands and results that mirror the original host-side call. Alternatively, if relying on a pre-existing library on the target, \lstIL{executor.call} can directly invoke an already-loaded symbol. Finally, \lstIL{executor.close} releases the session.

The remote execution is implemented using LLVM ORC v2~\cite{orcv2}. A \lstIL{catalyst-executor} process exposes the target process through ORC's Executor Process Control (EPC) mechanism. Catalyst emits an object file containing machine code for the selected target, and the host runtime transfers it to the executor, where ORC links and loads it into the target process. External symbols in the object are resolved at link time against shared libraries already loaded in the executor. The executor and its supporting runtime libraries are built separately for each target system, using a target sysroot when cross-compilation is required.

\subsubsection{The \lstIL{transport} dialect}
The transport dialect expresses the instructions that set up and drive the latency-critical
communication path. The dialect models a connection-oriented request/reply session between two endpoints and covers the whole life cycle, from creating the session and setting up the data path, to sending rounds and tearing down.

Sessions are typed by their role, as \lstIL{!transport.session<controller>} or \lstIL{!transport.session<coprocessor>}. To bring up a node's communication, \lstIL{transport.connect} establishes the connection and \lstIL{transport.exchange_keys} exchanges the memory-region information required by the transport. Each operation also has a non-blocking form (\lstIL{transport.connect_async}, \lstIL{transport.exchange_keys_async}) completed by \lstIL{transport.await}, allowing bring-up steps to overlap where the placement permits. \lstIL{transport.establish_channel} then arms the channel for the selected transport. Before rounds begin, \lstIL{transport.set_message_sizes} configures the request and reply sizes, and \lstIL{transport.start} readies the session. On the controller, each round resolves its session with \lstIL{transport.get_session}, stages the payload with \lstIL{transport.stage_payload}, and transmits it with \lstIL{transport.post}. The reply is received into a buffer with \lstIL{transport.collect}; where possible, \lstIL{transport.reply_slot} exposes the transport-owned reply buffer directly. On the coprocessor, \lstIL{transport.set_coprocessor_fn} binds the processing function used by the selected target; depending on the target device, this function may process individual messages or start a persistent service. Finally, \lstIL{transport.stop} and \lstIL{transport.destroy} tear down the session.

\begin{listing}
\begin{lstlisting}[language=MLIR, style=mlirstyle, escapeinside={(*}{*)}]
// Bring-up, emitted once into the program
// entry.
%ct = transport.create {
      backend_lib = "lib...cpu_verbs_controller.so",
      config = "dev=mlx5_1;gid=3",
      key = "gpu-coproc"}
      -> !transport.session<controller>
(*\step{1}*) transport.connect %ct {
        peer = "192.168.1.2",
        oob_port = 18560 : ui16}
      : !transport.session<controller>
(*\step{2}*) transport.exchange_keys %ct
      : !transport.session<controller>
(*\step{3}*) transport.establish_channel %ct "rdma"
      : !transport.session<controller>
(*\step{4}*) transport.set_message_sizes %ct {
        in_bytes = 8 : i64, out_bytes = 8 : i64}
      : !transport.session<controller>
(*\step{5}*) transport.start %ct
      : !transport.session<controller>

// One round, emitted in place of each decode.
%s = transport.get_session {key = "gpu-coproc"}
      : !transport.session<controller>
(*\step{6}*) transport.stage_payload %s, %syndrome {
        decoder_id = 0 : i32}
      : !transport.session<controller>,
        memref<3xi1>
(*\step{7}*) transport.post %s
      : !transport.session<controller>
%rep = transport.reply_slot %s
      : !transport.session<controller>
      -> memref<1xindex>
(*\step{8}*) transport.collect %s, %rep
      : !transport.session<controller>,
        memref<1xindex>
\end{lstlisting}
\caption{Transport operations in MLIR for session bring-up and one request/reply round, as the compiler emits them on the controller. Circled numbers mark the steps that also appear in Fig.~\ref{fig:backline_round}.}
\label{lst:transport_mlir}
\end{listing}

\subsubsection{Compilation flow and the QEC pipeline}\label{sec:compile_flow}
The frontend placement is serialized as an attribute on the generated MLIR module, from which every pass below reads the nodes, their targets, and the transport. A Backline placement augments Catalyst's base pipeline with transport passes; if it also requests implicit QEC, the QEC encoding and physical-lowering passes are added. Both compose with Catalyst's existing quantum and classical transformations. The relevant passes run in the following order:

\begin{itemize}
    \item With implicit QEC enabled, the QEC passes lower the program from a logical to a physical qubit representation. \lstIL{convert-quantum-to-qecl} raises the logical circuit into the \lstIL{qecl} dialect, \lstIL{inject-noise-to-qecl} inserts the noise model used by the demonstrations, and \lstIL{convert-qecl-to-qecp} followed by \lstIL{convert-qecp-to-quantum} expands each logical gate into its encoded physical circuit with a round of error correction around it. Each round emits a \lstIL{qecp.decode_esm_css} operation carrying the code's Tanner graph and the stabilizer type it decodes.
    \item \lstIL{inject-transport-session} reads the backline placement and inserts session creation and bring-up into compiler-generated setup functions, with the corresponding operations for teardown. For nodes assigned to executors, these life-cycle operations are placed in separately compiled components.
    \item \lstIL{lower-decode-to-transport} runs after bufferization and replaces each bufferized \lstIL{qecp.decode_esm_css} operation with a session lookup and one transport round --- \lstIL{stage_payload}, \lstIL{post}, and \lstIL{collect} --- taking the reply through \lstIL{reply_slot} where the target device exposes a transport-owned buffer.
     \item \lstIL{cross-compile-targets} then compiles each target component to an object file for its selected compilation target.
    \item \lstIL{dispatch-executor-targets} replaces calls to target components assigned to executors with an \lstIL{executor.open}/\lstIL{send_binary}/\lstIL{launch} sequence.
    \item In the final LLVM conversion stage, \lstIL{convert-executor-to-llvm} followed by
    \lstIL{convert-transport-to-llvm} lowers the two dialects to calls into their runtime C interfaces. When implicit QEC is enabled, \lstIL{convert-qecp-to-llvm} subsequently lowers any remaining QEC physical operations.
\end{itemize}

The implicit path currently supports the Steane code, selected with \lstIL{qec_code="steane"}. To use a different QEC code or to configure where in the circuit the rounds execute, the encoding and decoding can be explicitly written in the QNode with \lstIL{qp.backline.decode}, which emits the transport round directly and does not pass through \lstIL{lower-decode-to-transport}. This is the path the qLDPC demonstrations of Section~\ref{sec:v_decoding} take.

\subsection{The runtime layer}\label{sec:runtime}
The compiled program reaches hardware through a shared library interface with a well-defined C ABI. By decoupling the compiled program from the underlying implementation via this standard ABI, users and developers can easily plug in a custom runtime tailored to their own bespoke hardware.

\subsubsection{Session life cycle and memory}
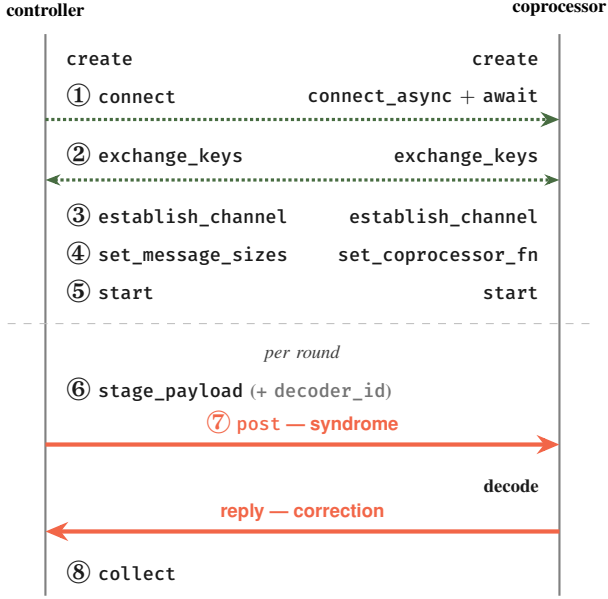
\begin{figure}[!htbp]
\centering
\begin{tikzpicture}[
  font=\sffamily\scriptsize\bfseries,
  life/.style={thick, black!50},
  msg/.style={-{Stealth[length=2.5mm, width=2mm]}, very thick, OliveGreen!70!black, densely dotted},
  data/.style={-{Stealth[length=3mm, width=2.5mm]}, ultra thick, Red!75},
  act/.style={black!90, font=\scriptsize\bfseries},
  note/.style={black!75, font=\scriptsize\bfseries\itshape},
]
  \def\cx{0}
  \def\dx{6.8}

  %% lifelines
  \node[font=\scriptsize\bfseries] (C) at (\cx,0) {controller};
  \node[font=\scriptsize\bfseries] (D) at (\dx,0) {coprocessor};
  \draw[life] (\cx,-0.32) -- (\cx,-7.75);
  \draw[life] (\dx,-0.32) -- (\dx,-7.75);

  %% ---------------- bring-up: once per session, out of band ----------------
  \node[act, anchor=west] at (\cx+0.15,-0.65) {\texttt{create}};
  \node[act, anchor=east] at (\dx-0.15,-0.65) {\texttt{create}};

  \node[act, anchor=west] at (\cx+0.15,-1.15) {\step{1}~\texttt{connect}};
  \node[act, anchor=east] at (\dx-0.15,-1.15)
    {\texttt{connect\_async} $+$ \texttt{await}};
  \draw[msg] (\cx,-1.45) -- (\dx,-1.45);

  \node[act, anchor=west] at (\cx+0.15,-1.95) {\step{2}~\texttt{exchange\_keys}};
  \node[act, anchor=east] at (\dx-0.15,-1.95) {\texttt{exchange\_keys}};
  \draw[msg, {Stealth[length=2mm]}-{Stealth[length=2mm]}] (\cx,-2.25) -- (\dx,-2.25);

  \node[act, anchor=west] at (\cx+0.15,-2.75) {\step{3}~\texttt{establish\_channel}};
  \node[act, anchor=east] at (\dx-0.15,-2.75) {\texttt{establish\_channel}};

  \node[act, anchor=west] at (\cx+0.15,-3.25) {\step{4}~\texttt{set\_message\_sizes}};
  \node[act, anchor=east] at (\dx-0.15,-3.25) {\texttt{set\_coprocessor\_fn}};

  \node[act, anchor=west] at (\cx+0.15,-3.75) {\step{5}~\texttt{start}};
  \node[act, anchor=east] at (\dx-0.15,-3.75) {\texttt{start}};

  %% ---------------- the round ----------------
  \draw[black!25, dashed] (-0.5,-4.15) -- (7.3,-4.15);
  % centred between the lifelines: anchored at the left edge it would have been
  % pierced by the controller lifeline at x=\cx, and moving it further left would
  % push the picture past \columnwidth.
  \node[note, font=\scriptsize\itshape] at ($(\cx,-4.55)!0.5!(\dx,-4.55)$) {per round};

  \node[act, anchor=west] at (\cx+0.15,-5.05)
    {\step{6}~\texttt{stage\_payload} \textcolor{black!55}{(+ \texttt{decoder\_id})}};
  \draw[data] (\cx,-5.75) -- (\dx,-5.75)
    node[midway, above] {\step{7}~\texttt{post} --- syndrome};
  \node[act, anchor=east] at (\dx-0.15,-6.30) {decode};
  \draw[data] (\dx,-6.90) -- (\cx,-6.90)
    node[midway, above] {reply --- correction};
  \node[act, anchor=west] at (\cx+0.15,-7.45) {\step{8}~\texttt{collect}};

\end{tikzpicture}
\caption{A representative remote setup and one decode round.
\lstIL{inject-transport-session} emits the one-time controller and coprocessor setup; dotted
arrows show the out-of-band interactions between them. Asynchronous operations used during
bring-up depend on the placement. \lstIL{lower-decode-to-transport} emits the solid request/reply round in place of each implicit decode. Circled numbers correspond to the controller-side operations in Listing~\ref{lst:transport_mlir}.}
\label{fig:backline_round}
\end{figure}

A target device keeps session state behind an opaque handle, and the interface is strictly ordered. For an RDMA-enabled target, each session goes through the following life cycle: first, a session is created from a named target library and role; \lstIL{connect} establishes the out-of-band handshake and initializes local endpoint state; \lstIL{exchange_keys} exchanges endpoint and memory-region metadata; \lstIL{establish_channel} configures data movement between the local and peer memory; then after role-specific configuration, \lstIL{start} readies the session for data exchange. After the required rounds, \lstIL{stop} and \lstIL{destroy} end the session. The details of memory provisioning and queue-pair state transitions depend on the target. Figure~\ref{fig:backline_round} shows the sequence of events within a pair of controller and coprocessor sessions. The memory is managed by the target device itself, and could allow for variable sizes and memory kinds, e.g. host RAM, HBM on GPUs, and DDR or BRAM on FPGAs. Regions are organized as ring buffers with a fixed slot count.

\subsubsection{Network transport and fabric selection}
Existing HPC-focused network-layer interfaces such as UCX~\cite{ucx2015} or the OpenFabrics Interfaces (OFI)~\cite{ofi2015} provide abstractions around the various low-latency RDMA layers, allowing one to express higher-level application logic without focus on the underlying network technologies.
Although these interfaces are generally preferred for their portability and have successfully supported applications across a wide range of HPC systems, their abstractions inevitably introduce some overhead. 
When performance is measured in nanoseconds to single-digit microseconds, it can be more effective to directly target the appropriate subsystems.

For the current state of this project, we opted for directly targeting and compiling against libibverbs~\cite{woodruff2005infiniband, kerr2011dissecting}, the user-space library that has become the common interface to the Linux RDMA subsystem~\cite{rdmacore} and associated tooling, originally developed around the InfiniBand specification~\cite{ibta2015infiniband}.
Given that many of the higher-level frameworks live above this layer (or related subsystems), potential overhead can be eliminated to achieve the fastest communication path. We anticipate this will have the widest impact, with the ability to adapt the associated tooling to work with other (and often more accessible) networking abstractions (e.g. UCX, OFI, ef\_vi, etc.) in follow-up work.

With this infrastructure, and to build the most widely compatible platform, we opted for RoCE~v2~\cite{roce2016}, which supports transmission using Ethernet packets via compatible RNICs and related FPGA infrastructure~\cite{ERNIC}. In addition, with the libibverbs layer, this allows us to immediately target any compatible RNIC via the upstream providers supported in the Linux RDMA subsystem, such as those from NVIDIA Mellanox (\lstIL{mlx5}), AMD Pensando (\lstIL{ionic}), Broadcom (\lstIL{bnxt\_re}), as well as many others, including the fallback software-defined emulation Soft-RoCE (\lstIL{rxe}) for testing. With this wide breadth of support, we can target any such provider as a testbed for building out quantum computing workloads across many provider platforms, as well as developing and testing locally using the emulation layer.

For the simplest prototyping use case, below even the emulation layer, the same session interface is implemented by a local target that replaces one-sided writes with in-process copies between the two roles' regions. This requires no verbs device at all, and is the mechanism behind the local development path of Section~\ref{sec:transport_mode}.

Besides portable networking abstraction libraries such as UCX or OFI, shared memory (SHMEM) libraries such as NVSHMEM and rocSHMEM are also highly relevant. These sit closer to the hardware and are tuned for low-latency RDMA communication in symmetric multi-device ecosystems (e.g. GPU-NIC-NIC-GPU). For systems that have baseline heterogeneity (CPU, GPU, RNIC), these libraries allow well-optimized application code to execute with best-in-class performance on the given systems. However, for more asymmetric designs, such as pairing a GPU and RNIC with an FPGA as an endpoint, many of these libraries are not immediately accessible, and require bespoke wrapping to achieve the lowest floor latency in communication. While it is possible to go below the libibverbs layer to directly tie into the RNIC drivers, such as using GPU-initiated RDMA writes, we found the CPU to be sufficient for our latency needs, with results reported in Section~\ref{sec:v_baseline}.

Furthermore, these libraries assume a performance model that is different from our quantum use-case. Indeed, systems built around these libraries achieve exceptional throughput and FLOPs by hiding latencies through the overlap of communications with compute, thereby minimizing upfront latency costs while maintaining high device utilization. This works for offline quantum workloads (e.g. pre-/post-processing), as well as ones with loose timing needs, but is of limited applicability for truly low-latency workloads, where device stalling while awaiting completion of a task can cause catastrophic build-up of corrections in the quantum system. In these cases, adding additional endpoints to our networking setup can be considered.

\subsubsection{Composing a system}\label{sec:composing}

It is worth stating the composability of these layers explicitly, since it is what makes one workload portable across different configurations. The transport, the controller platform, the coprocessor platform, their locality, and the definition of the coprocessor function are independent choices, summarized in Table~\ref{tab:backline_axes}. Moving a prototype from laptop onto HPC-grade hardware means changing entries within the table, without changing the workload.

In the following section we demonstrate a variety of setups, including local and remote endpoints, and controllers and coprocessors across multiple architectures, including the use of an FPGA as a controller. This is to emphasize that on a real quantum hardware system, the controller may not necessarily be a CPU host processor. Indeed, qubit control, measurement data acquisition, and the primary real-time decoding commonly reside on FPGAs, and ultimately on ASICs. This is because the real-time control tier (\ref{enum:realtime}) requires a deterministic sub-microsecond budget, which bare CPUs are unable to provide. The key claim that we seek to demonstrate with Backline is that the same Python program can scale from a fully local, single-process execution to an FPGA controller with a remote CPU or GPU coprocessor, changing only the placement while leaving the circuit untouched.

\begin{table}[!htbp]
\centering\footnotesize
\begin{tblr}{colspec={lX[l]},row{1}={font=\bfseries},rowsep=1.5pt}
\toprule
Axis & Choices \\
\midrule
Transport
  & memcpy -- in-process copies; no fabric hardware required \\
  & rdma -- RoCE~v2 over libibverbs, any supported RNIC provider \\
\midrule
Controller
  & CPU  -- Any CPU (for rdma, with connection to a NIC)\\
  & FPGA -- Application on APU, RDMA through e.g. Xilinx ERNIC engine \\
\midrule
Coprocessor
  & CPU -- per-message callback \\
  & GPU -- launch-once persistent kernel \\
\midrule
Coprocessor
  & C++ library -- most control, referenced by symbol \\
function
  & Triton kernel -- written in Python, AOT-compiled for the GPU \\
  & Tanner graph -- Hx/Hz as input, BP-OSD decoder generated \\
\midrule
Locality
  & Local in-process -- no executor \\
  & Local out-of-process -- a local executor \\
  & Remote machine -- an executor and a bundle deployed \\
\bottomrule
\end{tblr}
\caption{The configuration axes of a Backline placement. Each row can be varied independently without changing the QNode.}
\label{tab:backline_axes}
\end{table}

\begin{figure*}[!htbp]
\centering
\includegraphics[width=0.7\linewidth,clip]{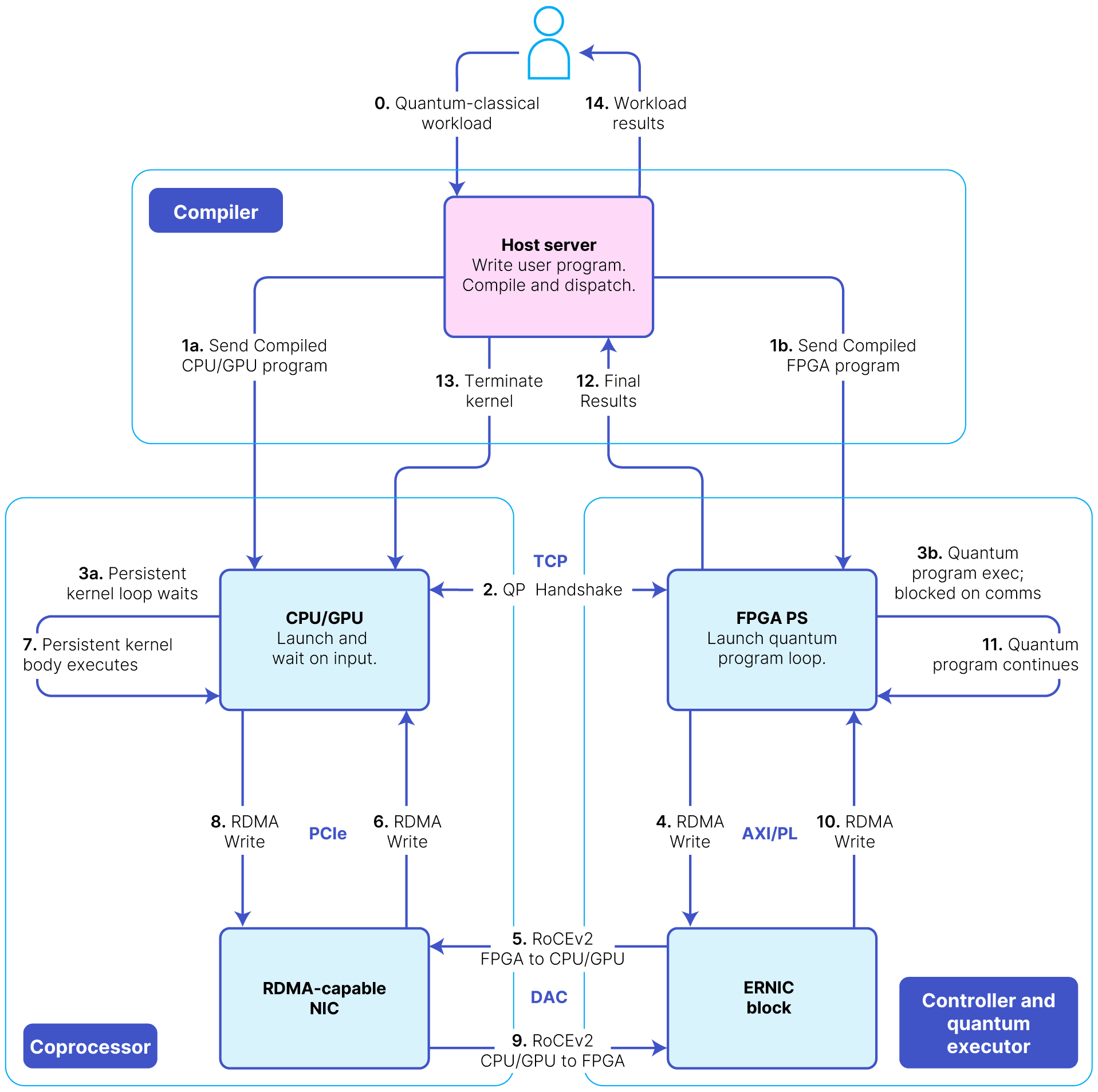}
\caption[]{Life cycle of a heterogeneous workload in Backline across a host, FPGA controller, and GPU coprocessor. Numbers indicate execution order. Steps 0-2 occur once per session. Steps 3-11 are one correction round and repeat for every decode. Steps 12-14 return results and tear down.}
\label{fig:sys_config}
\end{figure*}

Bringing everything together, an example workload's life cycle through our Backline design is outlined in Fig.~\ref{fig:sys_config}. 

\section{Demonstrations and evaluation}\label{sec:workloads}
The demonstrations and measurements in this section are all in the accompanying
repository\footnote{\url{https://github.com/PennyLaneAI/backline}}, which includes detailed installation and execution instructions, along with a cross-build system used for producing the stack deployed to each remote machine. The demonstrations characterize the performance of our benchmark setup while illustrating the breadth of the abstractions introduced in Section~\ref{sec:lowlatency}, spanning configurations from a laptop with no specialized hardware to an FPGA controller driving a GPU decoder across a fabric, with the decoder function either pre-compiled, written directly in Python, or generated from a code's parity-check matrices.

\subsection{Hardware setup}\label{sec:v_config}

The demonstrations are arranged in four hardware configurations with increasing setup requirements, allowing users to test progressively more capabilities. The first configuration only requires a single Linux CPU machine and nothing else. For demonstrations using local executors, the entire PennyLane and Catalyst program is compiled and executed on the local machine; for those using remote executors, the local machine instead handles compilation and orchestration, while the target components execute on one or more remote machines. The second configuration additionally requires a verbs device on the same machine; kernel Soft-RoCE is sufficient for this purpose, making an RDMA NIC optional. The third adds \textit{the server} with an RDMA NIC and a GPU. For our demonstrations, we relied on an AMD Threadripper PRO 9975WX CPU with an AMD Instinct MI210 GPU (PCIe~4.0) on an ASRock WRX90 WS EVO motherboard carrying a ConnectX-7 RNIC (PCIe~5.0). The fourth configuration adds an \textit{FPGA board}, an AMD Xilinx VPK120 board~\cite{VPK120} running the ERNIC IP block~\cite{ERNIC} with a customized reference design\footnote{Board images and setup instructions are available at \url{https://github.com/PennyLaneAI/backline-vpk120}.}, whose processing system is a pair of \texttt{aarch64} Arm Cortex-A72 cores. The \textit{FPGA board} is connected to \textit{the server} via the RDMA NIC using a 100~Gb Direct Attach Copper cable.

For the FPGA board running the controller node, we implemented a hardware-handshake (HWHS) engine in the programmable logic of the FPGA adjacent to the ERNIC RoCE engine. This allows us to avoid issuing work requests from the processing system that would place the application processing unit (APU), Linux scheduler, and repeated memory-mapped I/O (MMIO) transactions on the latency-critical path. Session setup remains a software control-plane operation, but the per-round operations of posting a work-queue entry (WQE), ringing the send-queue doorbell, detecting the reply, and measuring the round-trip latency are all performed by a finite-state machine inside the HWHS engine. 

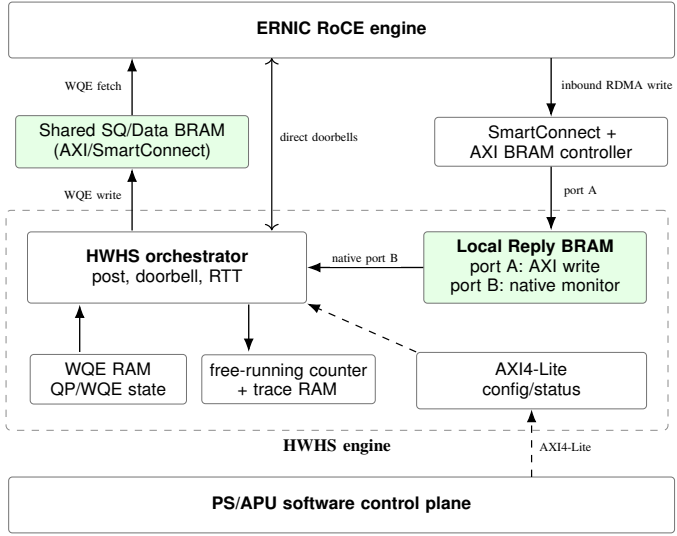
\begin{figure}[!htbp]
\centering
\resizebox{\columnwidth}{!}{%
\begin{tikzpicture}[
  font=\sffamily\scriptsize,
  block/.style={draw=black!55, rounded corners=1.5pt, fill=white,
    align=center, minimum height=7mm, text width=31mm},
  memory/.style={block, fill=green!10},
  datapath/.style={-Latex, line width=0.55pt},
  control/.style={-Latex, dashed, line width=0.5pt},
  edgelabel/.style={font=\tiny, align=center, fill=white, inner sep=1pt}
]
\node[block, text width=93mm, minimum height=8mm] (ernic) at (0,0)
  {\textbf{ERNIC RoCE engine}};
\node[memory] (shared) at (-30mm,-16mm)
  {Shared SQ/Data BRAM\\(AXI/SmartConnect)};
\node[block] (replyctrl) at (30mm,-16mm)
  {SmartConnect +\\AXI BRAM controller};
\node[block, text width=107pt, minimum height=29pt] (fsm) at (-25mm,-34mm)
  {\textbf{HWHS orchestrator}\\post, doorbell,
RTT};
\node[memory, text width=84pt] (replyram) at (2.77,-3.4)
  {\textbf{Local Reply BRAM}\\port A: AXI write\\port B: native monitor};
\node[block, text width=57pt] (state) at (-3.34,-5)
  {WQE RAM\\QP/WQE state};
\node[block, text width=64pt] (trace) at (-0.76,-5)
  {free-running counter\\+ trace RAM};
\node[block, text width=87pt, minimum height=23pt] (axil) at (2.72,-5.02)
  {AXI4-Lite\\config/status};
\node[draw=black!45, rounded corners=2pt, dashed,
  fit=(fsm)(replyram)(state)(trace)(axil), inner sep=3mm,
  label={[font=\scriptsize]below:\textbf{HWHS engine}}] (engine) {};
\node[block, text width=93mm, minimum height=8mm] (apu) at (0,-68mm)
  {\textbf{PS/APU software control plane}};

\draw[datapath] (shared.north) -- node[edgelabel, left=3pt]{WQE fetch} (-3,-0.4);
\draw[datapath] (-3,-2.87) -- node[edgelabel, left=3pt, pos=0.55]{WQE write} (shared.south);
\draw[datapath] (3,-0.4) -- node[edgelabel, right=3pt, pos=0.46]{inbound RDMA write} (replyctrl.north);
\draw[datapath] (replyctrl.south) -- node[edgelabel, right=4pt, pos=0.38]{port A} (3,-2.87);
\draw[datapath] (replyram.west) -- node[edgelabel, above]{native port B} (fsm.east);
\draw[datapath, <->, line width=0.5pt] (-1,-2.87) to[]
  node[edgelabel, pos=0.47, above right=3pt]{direct doorbells} (-1,-0.4);
\draw[datapath] (-3.75,-4.61) -- (-3.75,-3.93);
\draw[datapath] (-1.33,-3.93) -- (-1.33,-4.61);
\draw[control] (2.72,-6.4) -- node[edgelabel, right=2pt, pos=0.44]{AXI4-Lite} (axil.south);
\draw[control] (axil.north west) -- (fsm.south east);
\end{tikzpicture}%
}
\caption{Architecture of the VPK120 hardware-handshake engine, which connects
the APU control plane, ERNIC, local request/reply storage, and hardware
orchestrator.}
\label{fig:hwhs_architecture}
\end{figure}

Figure~\ref{fig:hwhs_architecture} gives an overview of the HWHS architecture and its connection to the ERNIC. The APU configures the engine through AXI4-Lite, while the HWHS orchestrator handles request posting, doorbells, reply
observation, and round-trip time (RTT) measurement in programmable logic. WQEs, queue-pair state, and reply data are kept close to the orchestrator so that repeated round-trip operations do not return to host software.

To reduce system-related overheads \textit{the server} treats each CPU chiplet as an individual NUMA
domain, via the NPS4 (NUMA Per Socket) BIOS option, and the RNIC and GPU are placed in PCIe
slots on the same domain so that traffic between them stays within one chiplet. The
controller process is pinned to a core on that domain and runs at real-time priority.

\subsection{Demonstrations}\label{sec:v_demos}

The demonstrations span four choices: where each node runs and on what kind of processor, which
transport carries a round between them, whether the error-correction round is inserted by the
compiler or written out by the user, and where the decode function comes from. Each of these choices is explained below; to illustrate, Listing~\ref{lst:v_reference} first shows our smallest demo in full: a logical GHZ state on
three qubits, corrected by a pre-compiled decoder in the same process. A few points worth noting:
\begin{enumerate}
    \item Both the controller and coprocessor are purely local, so no remote execution (or configuration) is required.
    \item \lstIL{qec\_code="steane"} turns the three logical gates into encoded circuits with a correction round around each, and each of those rounds into a message to the coprocessor.
    \item \lstIL{coprocessor_fn=steane_decode} picks a pre-compiled \lstIL{"steane_coprocessor"} function to perform the per-message decoding.
    \item \lstIL{transport="memcpy"} means syndrome and reply messages are transferred purely locally; no RDMA device is required.
\end{enumerate}

\begin{listing}
\begin{lstlisting}[language=Python]
import pennylane as qp

STEANE_CPU_DECODER_LIB_PATH = "/path/to/libsteane_coprocessor_cpu.so"

steane_decode = qp.CoprocessorFunction(
    name="steane_coprocessor", lib_path=STEANE_CPU_DECODER_LIB_PATH)

ctrl = qp.Controller(
    hardware="cpu",
    device=qp.device("null.qubit", wires=3))
coproc = qp.Coprocessor(
    hardware="cpu", coprocessor_fn=steane_decode)

dev = qp.Backline(controller=ctrl,
                  coprocessors=[coproc],
                  transport="memcpy",
                  qec_code="steane")

@qp.qjit(capture=True)
@qp.set_shots(10)
@qp.qnode(dev, mcm_method="one-shot")
def ghz():
    qp.Hadamard(0)
    qp.CNOT([0, 1])
    qp.CNOT([1, 2])
    return qp.sample([qp.measure(0),
                      qp.measure(1),
                      qp.measure(2)])

print("samples:", ghz())
\end{lstlisting}
\caption{The smallest complete program: local CPU to local CPU over the \lstIL{memcpy} transport. Asking for \lstIL{qec\_code="steane"} is what turns the three logical gates into encoded circuits with a correction round around each, and each of those rounds into a message to the coprocessor.}
\label{lst:v_reference}
\end{listing}

\subsubsection{Placement}\label{sec:v_placements}

As defined in Section~\ref{sec:backline_ui}, a Placement contains one controller, zero or more coprocessors, and the transport between them.

\medskip\noindent\textbf{The nodes.}
Across the demonstrations the controller is a local CPU, a remote CPU, or a remote FPGA, and
the coprocessor is a local CPU or a remote GPU. Listing~\ref{lst:v_placement} shows four of the placements we showcase in our demos. 

\begin{listing}
\begin{lstlisting}[language=Python]
# demo 1  -- both roles are local
ctrl = qp.Controller(
    hardware="cpu",
    device=qp.device("null.qubit", wires=3))
coproc = qp.Coprocessor(
    hardware="cpu", coprocessor_fn=steane_decode)
dev = qp.Backline(controller=ctrl,
    coprocessors=[coproc], transport="memcpy",
    qec_code="steane")

# demo 1a -- the same pair, over a verbs device
ctrl = qp.Controller(
    hardware="cpu",
    device=qp.device("null.qubit", wires=3),
    init_args={...})
coproc = qp.Coprocessor(
    hardware="cpu", coprocessor_fn=steane_decode,
    endpoint=qp.Endpoint(...),    # soft-RoCE
    init_args={...})
dev = qp.Backline(controller=ctrl,
    coprocessors=[coproc], transport="rdma",
    qec_code="steane")

# demo 3  -- remote CPU controller, remote GPU
ctrl = qp.Controller(
    hardware="cpu", remote=True,
    executor_options={...},
    init_args={...})
coproc = qp.Coprocessor(
    hardware="gpu", remote=True,
    coprocessor_fn="gpu_steane_launcher",
    endpoint=qp.Endpoint(...),    # fabric address
    executor_options={...},
    init_args={...})
dev = qp.Backline(controller=ctrl,
    coprocessors=[coproc], transport="rdma",
    qec_code="steane")

# demo 4  -- the controller moves to the FPGA;
#            everything else is demo 3 unchanged
ctrl = qp.Controller(
    hardware="fpga", remote=True,
    executor_options={...},
    init_args={...})
\end{lstlisting}
\caption{Four placements for the same program. Demo~1 and demo~1a differ in the transport alone; demo~1a to demo~3 changes the machine and the device class together; demo~3 to demo~4 changes only the controller. The \lstIL{qp.Backline} line is the placement itself, carrying the two nodes and the transport between them. The ellipses in the constructors abbreviate testbed-specific values declared once in the repository's \lstIL{placement.py} for easy definition and re-use, which are shared by every demonstration.}
\label{lst:v_placement}
\end{listing}

A node is characterized by the machine its compiled code runs on and what it loads there. The following are arguments used to configure a node: \lstIL{remote} decides whether the code is shipped instead of run in-process, \lstIL{executor\_options} describes how to reach the machine over a network, \lstIL{endpoint} specifies the coprocessor's address on the RDMA fabric, \lstIL{init_args} carries target-specific setup, and \lstIL{hardware} is paired with the placement's transport to resolve a concrete target device library.

\medskip\noindent\textbf{The transport.}\label{sec:v_transport}
The two transports of Section~\ref{sec:transport_mode} divide the demonstrations. Demo~1 uses
\lstIL{memcpy}, which needs no verbs device at all and is the one to run first; every other
demonstration uses \lstIL{rdma}. Demo~1a is the useful step between them, being demo~1 over a
Soft-RoCE device: the verbs path is genuine while the transport beneath it is still software, so
it establishes that the path works. Given a real RDMA NIC, the same
placement runs over that NIC's loopback on one machine, or over a wire between two.

\subsubsection{Implicit and explicit decoding}\label{sec:v_decoding}

With \lstIL{qec\_code="steane"} given on the device, the user defines only a \emph{logical
circuit}, and points the compiler at the passes and libraries that do the rest: error correction
is applied automatically within the stack, with the encoding introduced at compile time by the QEC lowering passes and the decoding performed on the coprocessor. Each logical gate is expanded into its physical circuit with a round of error correction around it --- extract the stabilizers, decode the syndrome, apply the correction --- none of which the user needs to specify.

Alternatively, users can express these steps explicitly in their Python program. The user can encode the logical
circuit by hand, in effect writing the physical circuit to be compiled and executed through
Backline, then measure the syndromes, decode them, and apply the correction explicitly. An explicit call to
\lstIL{qp.backline.decode} sends a syndrome to the
coprocessor and blocks until the correction comes back. Where several decoders are registered, a
\lstIL{decoder\_id} (which is part of the payload) selects between them.

\lstIL{decode} expands into a series of local runtime calls in the controller's process. The same
operations are accessible directly through the \lstIL{qp.runtime\_call} interface of Section~\ref{sec:backline_ui}: \lstIL{get\_session} finds the controller's open
session with the named coprocessor, \lstIL{stage\_payload} writes the packed syndrome into the
outgoing slot, \lstIL{post} hands that slot to the transport, and \lstIL{collect} blocks until
the correction lands in the reply slot. Demo 2a reproduces demo 2's result with the
exchange written out this way. The three levels differ only
in how much is left to the compiler.

\begin{listing}
\begin{lstlisting}[language=Python]
# (a) implicit - demos 1, 1a, 3, 4, 5
#     a logical circuit; the compiler inserts
#     the encoding and the decode round
dev = qp.Backline(..., qec_code="steane")

# (b) explicit - demo 2
z_syn, x_syn = extract_syndromes()
corr_z = qp.backline.decode(x_syn, decoder_id=0)
corr_x = qp.backline.decode(z_syn, decoder_id=1)
apply_correction(corr_x, qp.X)
apply_correction(corr_z, qp.Z)

# (c) runtime calls - demo 2a
#     what (b) is made of
session = qp.runtime_call(
    GET_SESSION, ROLE_CONTROLLER, "gpu-coproc")
qp.runtime_call(STAGE_PAYLOAD, session, packed,
                PACKED_BYTES, decoder_id)
qp.runtime_call(POST, session, WORK_ITEM)
_status, correction = qp.runtime_call(
    COLLECT, session, PACKED_BYTES,
    out_bytes=PACKED_BYTES)
\end{lstlisting}
\caption{The same round at three levels. In (b) the syndrome is extracted by the user and the correction applied to live qubits, with \lstIL{decoder\_id} choosing which of the coprocessor's decoders serves it. In (c) that one call is replaced by the four it expands to.}
\label{lst:v_decoding}
\end{listing}

\subsubsection{The decoder function}\label{sec:v_decoder_fn}

The demonstrations exercise all three routes of Section~\ref{sec:coproc_fn}: a pre-compiled decoder for the
$[[7, 1, 3]]$ Steane code, used on both a CPU and a GPU coprocessor; a Steane decoder written in Python as a
Triton kernel; and a belief-propagation decoder with ordered-statistics post-processing~\cite{Roffe:2020efe}
generated from the parity-check matrices of the $[[13, 1, 3]]$ hypergraph-product code, run for 10 iterations.
Several decoders may be registered on one coprocessor and selected per round by the \lstIL{decoder\_id}
travelling with the payload, which is how a single coprocessor serves the $X$ and $Z$ stabilizer types with a
decoder each.

Table~\ref{tab:v_demos} lists the demonstrations available in the repository, their respective choices on the
above, and the route each one takes to its decoder.

\begin{table}[!htbp]
\centering\footnotesize
\begin{tblr}{colspec={llllll},row{1}={font=\bfseries},rowsep=1.2pt,colsep=3pt}
\toprule
& Controller & Coprocessor & Transport & Decode & Decode fn \\
\midrule
1  & local CPU   & local CPU  & memcpy & implicit & pre-compiled \\
1a & local CPU   & local CPU  & rdma\TblrNote{a} & implicit & pre-compiled \\
2  & remote CPU  & remote GPU & rdma & explicit & Triton BP-OSD \\
2a & remote CPU  & remote GPU & rdma & runtime  & Triton BP-OSD \\
3  & remote CPU  & remote GPU & rdma & implicit & pre-compiled \\
4  & remote FPGA & remote GPU & rdma & implicit & pre-compiled \\
5  & remote FPGA & remote GPU & rdma & implicit & Triton Steane \\
\bottomrule
\end{tblr}
\caption{All demonstrations included in the repository, as combinations of the choices of
Sections~\ref{sec:v_placements} to \ref{sec:v_decoder_fn}. \emph{Decode} is how the round is
expressed: inserted by the compiler (implicit), written with \lstIL{qp.backline.decode} (explicit), or written out
as runtime calls (runtime). \emph{Decode fn} is where the decoder comes from: a symbol in a pre-compiled CPU or GPU
library, generated from the code's parity-check matrices (Triton BP-OSD), or directly through a Triton kernel (Triton Steane).
\TblrNote{a}~over kernel Soft-RoCE, so no RDMA NIC is required.}
\label{tab:v_demos}
\end{table}

\subsection{Measurements}\label{sec:v_measurements}

\subsubsection{Baseline}\label{sec:v_baseline}

To accurately measure the overheads, we built an end-to-end latency evaluation benchmark (\lstIL{bench.py}) to determine the floor latency of the ecosystem using the hardware configuration specified in Section~\ref{sec:v_config}. For the baseline case, the FPGA issues a 16-byte payload (including an 8-byte syndrome) as a one-sided RDMA write to a given target (CPU or GPU memory). Through the hardware-handshake engine, the programmable logic (PL) directly posts that write and detects the reply, so there is no host involvement in the round trip. To ensure a fast response, the target coprocessor polls on the expected memory buffer, then in the CPU case echoes the reply back directly through a one-sided write to the FPGA, or in the GPU case signals the CPU, which then issues the one-sided write back. The FPGA controller times the round trip in its own clock domain.

\begin{figure}[!htbp]
\centering
\begin{subfigure}{\columnwidth}
  \centering
  \includegraphics[width=\linewidth]{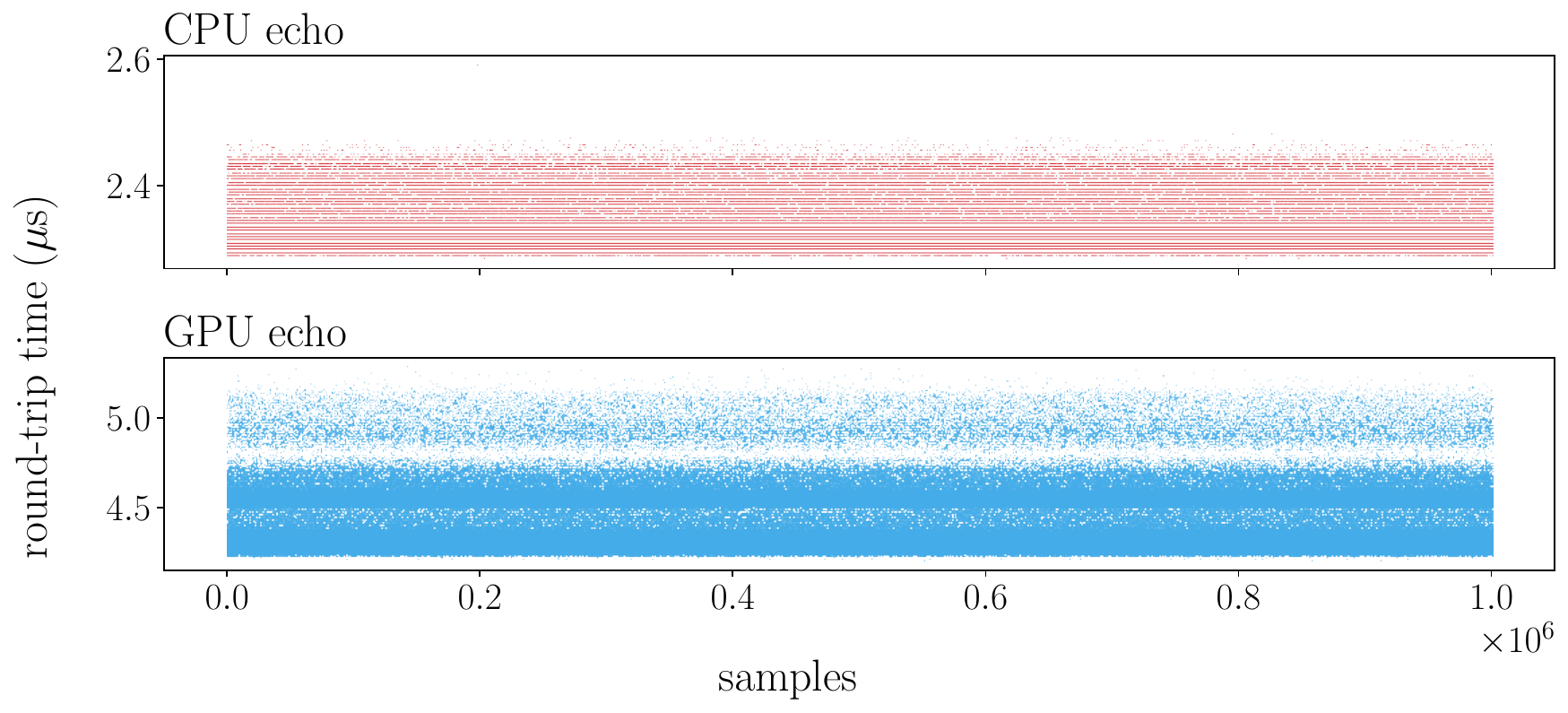}
  \caption{Round-trip time against sample number.}
  \label{fig:v_baseline_scatter}
\end{subfigure}
\par\medskip
\begin{subfigure}{\columnwidth}
  \centering
  \includegraphics[width=\linewidth]{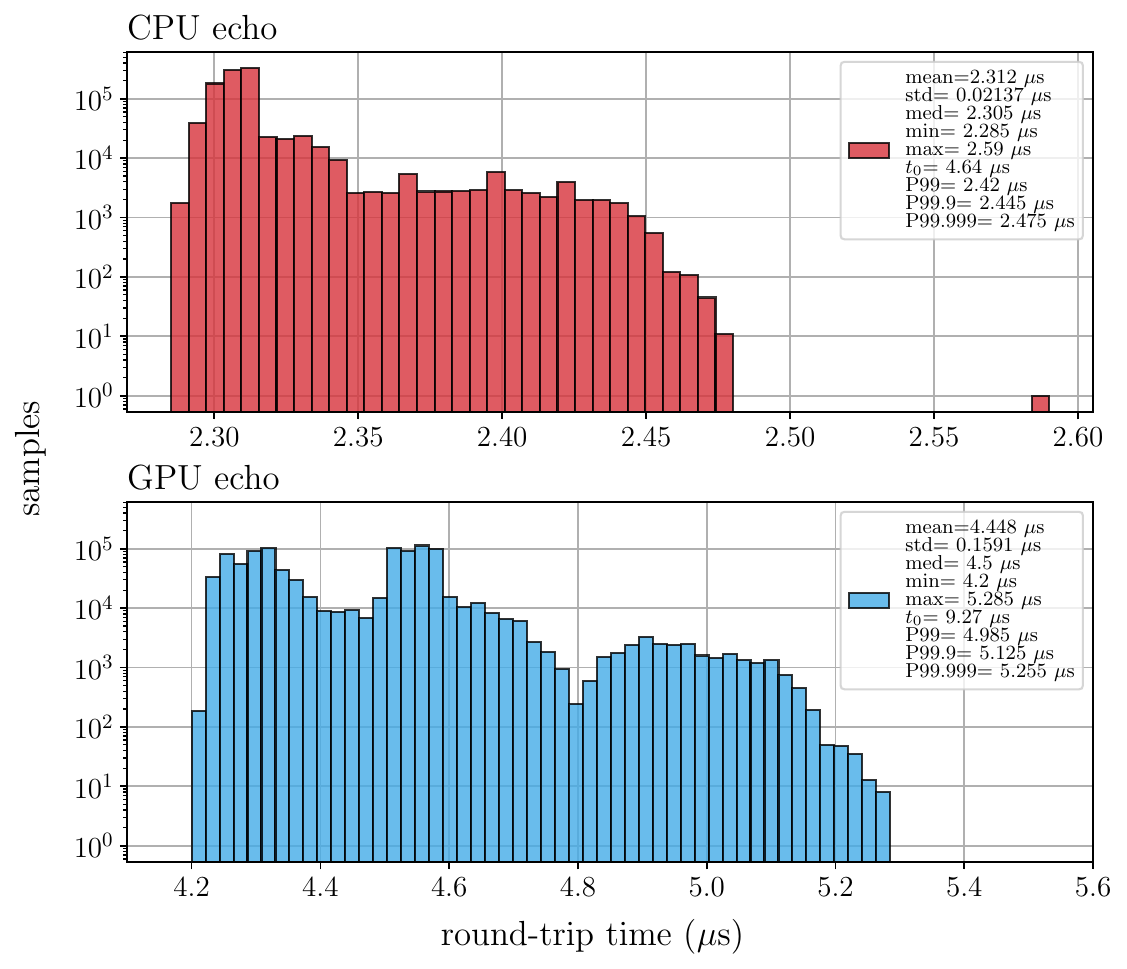}
  \caption{Round-trip time distribution.}
  \label{fig:v_baseline_hist}
\end{subfigure}
\caption{Baseline steady-state round-trip latency over $10^6 -1$ rounds from FPGA controller to and from CPU or GPU coprocessor. Timing is measured by the FPGA clock domain, with the first warm-up sample excluded from all plots. Statistical quantities are given in the legend of each distribution plot, with the excluded warm-up value given by $t_0$.}
\label{fig:v_baseline}
\end{figure}

Figure~\ref{fig:v_baseline} shows the timing and distribution of $10^6-1$ round trips initiated from
the FPGA controller to and from a CPU and GPU coprocessor device, as a time series and as a
distribution. Both panels show the steady state, with a single initial warm-up round trip
excluded.\footnote{The first warm-up round trip costs \qty{4.64}{\us} on the CPU path and
\qty{9.27}{\us} on the GPU path, in both cases far outside the steady-state distribution that follows it. Full results are shown in Appendix~\ref{test-environment}}. From the data, the CPU path produced a median of \qty{2.305}{\us}, a mean of \qty{2.312}{\us}, a standard deviation of \qty{21.4}{\ns}, and a P99.999 tail at \qty{2.475}{\us}; the GPU path measured a median of \qty{4.5}{\us}, a mean of \qty{4.448}{\us}, a standard deviation of \qty{0.159}{\us}, and a P99.999 of \qty{5.255}{\us}. A full description of the FPGA timing setup and further experiments is given in Appendix~\ref{test-environment}.

The baseline measurements suggest that this framework is suitable for the synchronous co-processing tier described in Section~\ref{section:qec} with room to spare. In particular, the CPU path is not only \qty{2.2}{\us} faster in the median but also over seven times tighter in the spread. Against a budget that has to be met on every round rather than on average, this spread in particular is decisive in whether a decoder is admissible, and on that measure the commodity CPU path is the stronger of the two.

\subsubsection{Decoder runtime}\label{sec:v_decoder_cost}

Using the same benchmarking script and setup, we measure the end-to-end round-trip timing from FPGA controller to a CPU or GPU coprocessor device including a decoder execution. Table~\ref{tab:v_responders} summarizes the time it takes for the three decoders of Section~\ref{sec:v_decoder_fn}, alongside an `echo' row giving the baseline of Section~\ref{sec:v_baseline} on the same path. The BP-OSD decoder is a naive implementation; there are many optimizations possible, including using hardware-friendly normalized min-sum to approximate more expensive operations on GPUs, which could be implemented directly within the Triton kernel. The point of the Backline platform is to enable quick turnarounds for such experimentation.

\begin{table}[!htbp]
\centering\footnotesize
\begin{tblr}{
  width=\columnwidth,
  colspec={lX[l]rrrrr},
  row{1}={font=\bfseries},
  rowsep=1.5pt, colsep=3pt
}
\toprule
Device & Coprocessor function & Mean & Median & P99 & P99.9 & Max \\
\midrule
CPU & Echo                &  2.312 &  2.305 &  2.420 &  2.445 &  2.590 \\
CPU & Steane decode       &  2.311 &  2.305 &  2.410 &  2.445 &  2.545 \\
\midrule
GPU & Echo                &  4.448 &  4.500 &  4.985 &  5.125 &  5.285 \\
GPU & Steane decode       &  4.412 &  4.405 &  4.945 &  5.105 &  5.265 \\
GPU & qLDPC BP-OSD decode & 31.109 & 31.075 & 32.410 & 32.680 & 33.260 \\
\bottomrule
\end{tblr}
\caption{Steady-state software-loop RTT statistics in microseconds. Round~0 is excluded.}
\label{tab:v_responders}
\end{table}

From the results, we observe that for very simple decoders, the decoding time is virtually free compared to the communication overheads. For more complex decoders, the decoder may take longer than the transport overhead. For prototyping and fast R\&D, defining the decoder directly in Python through Triton provides the least resistance, but for best performance, users should write a tuned and optimized version directly as a CPU function or GPU kernel. Backline is designed to facilitate both modes, and easy transition from one to the other.

\section{Conclusions and outlook}\label{sec:conclusions}

\begin{figure*}[!htbp]
\centering
\includegraphics[width=0.85\linewidth,clip]{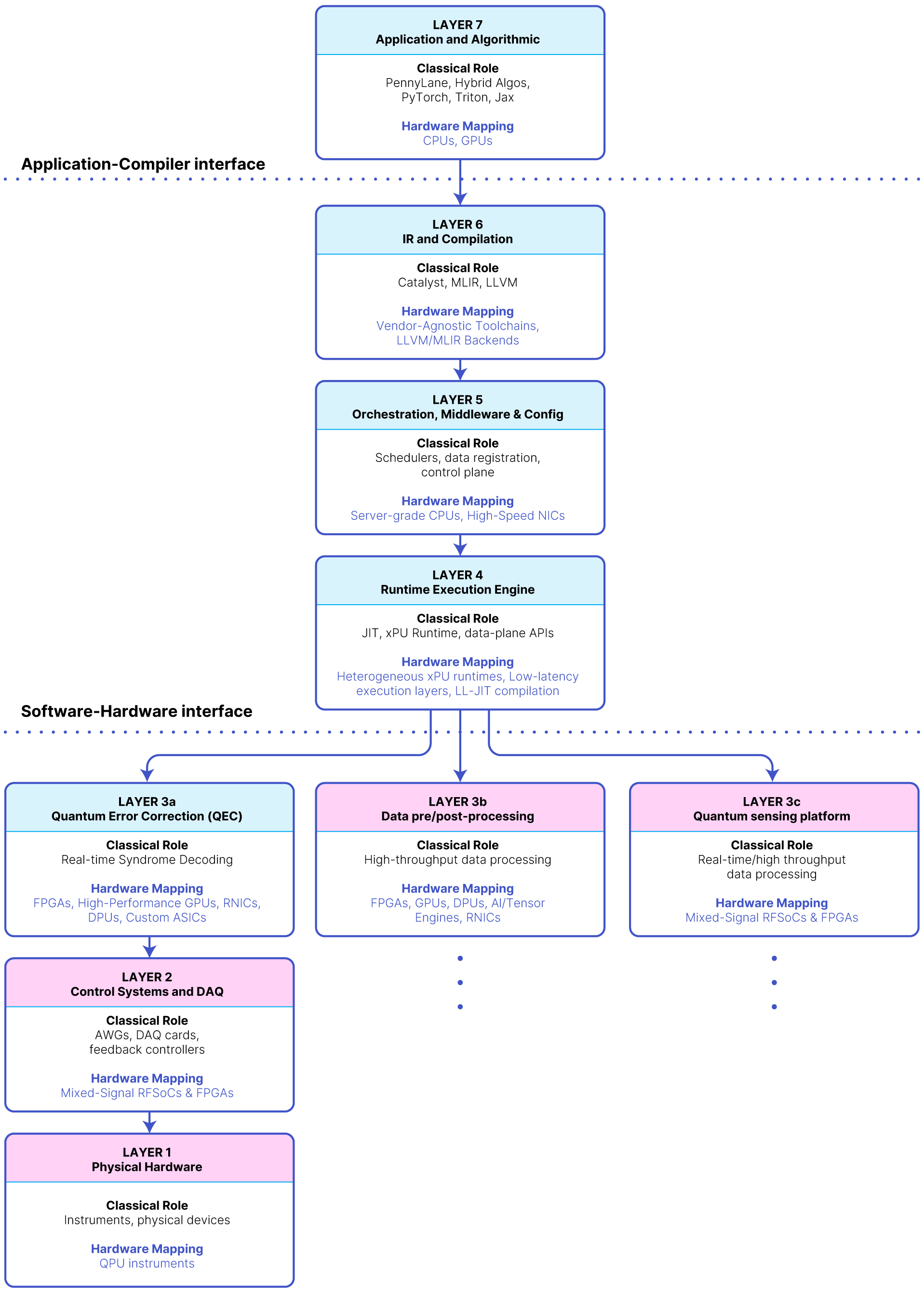}
\captionsetup{width=0.85\textwidth}
\caption[]{An open architecture pathway for FTQC and adjacent quantum systems. Blue denotes what Backline
supports today: the application and compilation layers it is built into, the orchestration and
runtime layers it supplies, and real-time QEC as the workload driving them. Pink denotes next
steps: high-throughput data pre- and post-processing, quantum sensing platforms, and reaching
further down through control and data acquisition to the physical instruments.}
\label{fig:next_step}
\end{figure*}

A quantum computer is not purely quantum. A fault-tolerant quantum computer requires an amalgamation of and complex interaction between quantum and classical systems, which encode, decode, measure, and react on the timescale of the quantum hardware itself. The difficulty of building and programming such systems is now a first-order constraint on the field. The solution to this is not to harden a single fast path, but to give researchers a way to express a workload once and place it across whichever classical devices the problem requires and that are available to them.

Backline is our answer to that. It adds a heterogeneous compilation and runtime layer to PennyLane and Catalyst that spans the required stack: a declarative frontend, in which a placement is defined by a controller, its coprocessors, and the transport between them; an operational compilation layer, in which the \lstIL{executor} and \lstIL{transport} dialects turn that placement into explicit sessions, staging and launching operations as well as lowering them alongside the other QEC passes in MLIR; and a runtime layer, whose unified ABI makes swapping target devices and transport modes seamless. Section~\ref{sec:workloads} exercised these layers against real hardware. The same GHZ program ran on a single workstation over \lstIL{memcpy}, on that workstation over \lstIL{rdma}, across a remote CPU
controller and GPU coprocessor, and finally from a Xilinx VPK120 controller to that same
GPU, with the differences between configurations confined solely to the placement.

The baseline measurements establish the achievable latency regime: driven from the FPGA controller under the hardware-handshake engine, the steady-state round trip to a CPU coprocessor measures a median of \qty{2.305}{\us} and to a
GPU coprocessor \qty{4.5}{\us} over $10^6-1$ rounds (Fig.~\ref{fig:v_baseline}). Both sit inside
the synchronous co-processing tier of Section~\ref{section:qec}, which is the regime that backup
decoding and auxiliary real-time processing occupy.

\subsection{Choosing a coprocessor}
An enterprise-grade GPU is not necessarily a prerequisite for real-time QEC. From our measurements, the CPU path is not only faster, but substantially more predictable. Against a deterministic budget, the tail is a crucial factor in deciding whether an external decoder is admissible, and on that measure the CPU path is stronger by a wide margin. This also lowers the hardware cost of entry for users who want to begin prototyping. 

This is not an argument that GPUs are unnecessary; the device should follow what the workload requires. Where the decode is cheap relative to the round trip, a CPU coprocessor is the better endpoint, since the GPU would be paying a transport penalty for parallelism it does not need. Where the decode itself dominates (e.g. for larger decoders, batched processing, or neural-network based decoders), the transport difference stops being the deciding term and the GPU's throughput may be what matters. Backline enforces neither choice, leaving it to the user to try both.

\subsection{Fabric is a moving target}
RoCE v2 gave us a latency floor in the single-digit microseconds, and that floor is what puts the
synchronous co-processing tier within reach. Similar experiments~\cite{lao2026realtimedecodingquantumerror} with other HPC-grade networking also establish a comparable floor. This is just the starting point. Emerging interconnects are converging on improved performance in every direction, for example UALink~\cite{ualink} for scaling up accelerator-to-accelerator links within a rack, and Ultra Ethernet~\cite{ultraethernet} for scaling out fabrics with transport optimizations surrounding both small latency-sensitive messages and bulk data throughput. We expect these to be important for FTQC infrastructure, enabling both high performance and flexible topologies.

The abstractions in Section~\ref{sec:lowlatency} were designed with exactly this flexibility in mind. The \lstIL{transport} dialect describes only the life cycle of a session; a new fabric is a new runtime target behind the same C ABI, and reaches the user as a new value for a single \lstIL{transport=} keyword. 

Currently the dialects and runtimes do not abstract over the message sizes. The message size is fixed, and every round in the session uses those exact byte counts. This is sufficient for the workloads shown here today, where the syndrome has a width known at compile time and a fixed slot allows registering the memory once and staging it, without allocation on the critical path. However, this may be unsuitable for cases where the payload varies from round to round, such as a decoder that returns a variable-length correction, or a controller batching a number of rounds that is only known at execution time. The current limitation of the transport layer is therefore support for variable payload sizes without giving up the pre-registered, allocation-free fast path; we leave this to future work.

\subsection{Scaling up}
With hardware system scale-ups happening to accommodate the increasing demand for AI, quantum software layers can
likely benefit from integration with such systems. Rack-scale designs such as AMD Helios (MI455X
GPUs, EPYC CPUs, Pensando RNICs) should be implicitly compatible with the Backline programming
abstractions, since a rack of accelerators behind an RDMA-capable NIC is the same object that our
placement model already describes, only with more coprocessors in it. This admits direct
multi-device targeting and compilation from a single program. We see this being beneficial not just
for simulation, but for full-system emulation of FTQC control stacks whose hardware designs are
approaching reality.

\subsection{Scaling down}
Access to enterprise-grade hardware is frequently the binding constraint for a research group, and
local development is what makes iteration on workloads, applications, and SDK layers fast. We deliberately kept a continuous path from a laptop to the full system. The same program runs over
\lstIL{memcpy} on one machine, over Soft-RoCE~\cite{rdmaRH9} against a software RDMA device with no
NIC present at all, over verbs on a workstation with a commodity RNIC, or over an FPGA's hardware
engine; moving between these requires only a change to the placement rather than to the workload.

Another part of enabling this accessible approach is keeping the framework open, Python-native, and infrastructure-agnostic. Backline is built and shipped alongside the open-source PennyLane and Catalyst codebases, with the frontend API, the MLIR dialects, and the runtime targets all available and extensible. We consider the hardware side to be extensible on the same terms: there is no restriction on the infrastructure or vendors for any of the CPU, GPU, and FPGA components, while open-source RNIC designs~\cite{balboa, coyote_v2} make it possible for research groups to implement and test different FPGA setups. Prototyping and co-designing QEC codes, decoders, and transport is entirely possible within Backline.

\subsection{Down to silicon}
The real-time control tier of Section~\ref{section:qec} does not end at the FPGA. As we noted in Section~\ref{sec:composing}, qubit control, data acquisition, and the primary decoder sit on FPGAs and ultimately on ASICs, since this tier requires a bounded, sub-microsecond worst-case latency that general-purpose hosts cannot provide. Figure~\ref{fig:next_step} summarizes in blue what Backline enables today and, in pink, the remaining steps needed to cover the whole quantum system in the near future.

Platforms for QEC infrastructure remain highly bespoke, and as we argued at the outset the field is
unlikely to be served by freezing a standard around today's hardware. What it does need is vendor-agnostic options, and interfaces that evolve with the field while remaining stable enough
to build against. Backline is a stride in that direction: an architecture for 
workloads spanning CPUs, GPUs, and FPGAs, with transport abstractions exposed at compile time and implemented under a runtime interface, reached from a Python frontend that quantum researchers already use.
Getting a decoder from a Python prototype onto real low-latency hardware should be a matter of
changing where it runs, with little or no need to rewrite the program. We envision Backline as a platform for researchers and engineers across the quantum computing stack. Quantum architecture and algorithm researchers can explore new designs for applications and error correction with minimal friction. Quantum hardware developers can use a cohesive, widely adopted environment to prototype control systems. Systems engineers and classical computing vendors can rely on a user-friendly abstraction to seamlessly orchestrate heterogeneous computing environments.

\section*{Acknowledgements}
We acknowledge support from AMD for providing GPU hardware to support this project. Additionally, we thank Jose M. Monsalve-Diaz, Paul Hartke, Yasuko Eckert, and Muhammad Osama for their discussions on the software side, as well as Rip Sohan, Rowan Lyons, Chris Neely, and Mike Crowley for assistance on the hardware side.
We also acknowledge significant help from the Xanadu software team members Ali Asadi and David Ittah for software contributions, discussions and reviews, Jason Selby and Isaac De Vlugt for manuscript review, as well as IT team members Nabil Dib, Andy Yen, Andrew Hodder, Sanchit Bapat, Seun Shoga, and Raymond Baksh for infrastructure setup and configurations.

\bibliography{main}

\appendices

\raggedbottom
\section{Test Environment and FPGA experiments}\label{test-environment}
\subsection{Reference test environment}
\subsubsection{CPU/GPU server}

The primary server used for our demonstrations and benchmarks is described in Table~\ref{tab:test_environment_server}. The GPU and RNIC occupy the same NUMA domain to avoid an inter-domain hop on the data path. The executor and its memory allocations are bound to that domain with \texttt{numactl}; the CPU polling thread is pinned to a fixed core and requests real-time scheduling.

\begin{table}[!htbp]
\centering
\footnotesize
\caption{Reference server used for demonstrations and benchmarking.}
\label{tab:test_environment_server}
\begin{tblr}{width=\columnwidth,colspec={l X[l]},rowsep=1.5pt}
\toprule
Component & Reference configuration \\
\midrule
Platform & ASRock WRX90 workstation \\
CPU & AMD Ryzen Threadripper PRO 9975WX, 32 cores / 64 threads \\
GPU & AMD Instinct MI210, CDNA2 / \texttt{gfx90a}, PCIe~4.0 \\
RNIC & NVIDIA ConnectX-7 / \texttt{mlx5}, PCIe~5.0, RoCE~v2 \\
NUMA layout & NPS4; RNIC, GPU, process, and buffers on one domain \\
\bottomrule
\end{tblr}
\end{table}

For benchmarking, we use the FPGA as controller, and the CPU or GPU on this server as the coprocessor to measure the round-trip timing. In the GPU case, the GPU executes a persistent kernel to poll on its memory for message arrival, then writes the response message to host memory; a separate CPU thread polls the host memory and then writes the message back to the controller. The CPU and GPU \texttt{echo} responders return the received value without decoding and therefore allow us to measure the transport floor. The \texttt{cpu-steane} and \texttt{gpu-steane} responders use pre-compiled Steane lookup decoder implementations. The \texttt{gpu-qldpc} responder runs a Triton-generated BP-OSD decoder on the MI210 GPU.

\subsubsection{FPGA controller}

The controller for our benchmarks is described in Table~\ref{tab:test_environment_vpk_hw}. Both the ARM processing system (PS) and the programmable logic (PL) participate in bringing up a session, but their responsibilities are separated, as summarized in Table~\ref{tab:test_environment_vpk}.

\begin{table}[!htbp]
\centering
\footnotesize
\caption{Reference FPGA controller used for demonstrations and benchmarking.}
\label{tab:test_environment_vpk_hw}
\begin{tblr}{width=\columnwidth,colspec={l X[l]},rowsep=1.5pt}
\toprule
Component & Reference configuration \\
\midrule
Platform & AMD VPK120 evaluation board \\
Device & AMD Versal Premium \texttt{XCVP1202} \\
Processing system & Arm (aarch64), PetaLinux 2024.2 \\
RDMA engine & AMD (Xilinx) ERNIC, RoCE~v2 \\
HWHS clock & \qty{200}{\mega\hertz}; one recorded cycle is \qty{5}{\nano\second} \\
\bottomrule
\end{tblr}
\end{table}

\begin{table}[!htbp]
\centering
\footnotesize
\caption{VPK120 controller partition in the current design.}
\label{tab:test_environment_vpk}
\begin{tblr}{width=\columnwidth,colspec={l X[l]},rowsep=1.5pt}
\toprule
Region & Responsibility \\
\midrule
PS & PetaLinux, session setup, QP/MR creation, configuration, readout \\
ERNIC PL & WQE consumption, doorbells, RoCE~v2 transmit/receive data path \\
HWHS PL & Per-round sequencing, WQE construction, reply detection, RTT timestamping \\
On-chip memory & SQ/data BRAM, dual-port reply URAM, RTT trace RAM \\
\bottomrule
\end{tblr}
\end{table}

\subsection{ERNIC and HWHS implementation}

\subsubsection{ERNIC software and hardware split} The Embedded RDMA Enabled NIC (ERNIC) is an AMD soft-core IP that implements FPGA-side transport and packet processing for RDMA over RoCE~v2 in programmable logic on an FPGA. The PS is responsible for host and device memory allocation, establishing and programming queue-pair (QP) contexts, and handling error exceptions. Hardware then implements the zero-copy data transport without software intervention. The ERNIC Core manages packet encapsulation and decapsulation, sequence validation, Direct Memory Access, and hardware acknowledgements/re-transmissions. Application-layer logic handling timing and dispatch are handled by the Hardware-Handshake (HWHS) module's orchestrator (\texttt{hh\_orchestrator}), which interfaces directly with the ERNIC's payload buses and doorbell registers.

\subsubsection{HWHS state machine}
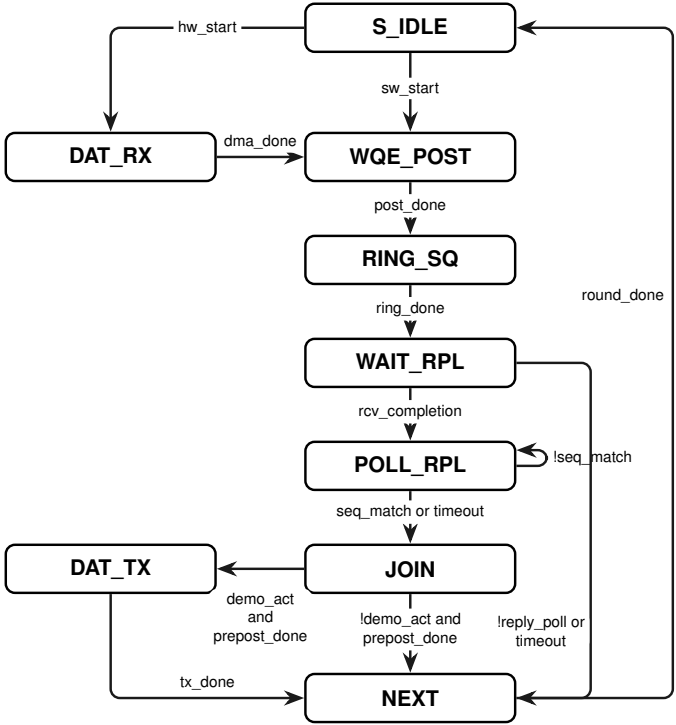
\begin{figure}[!htbp]
\centering
\resizebox{\columnwidth}{!}{%
\begin{tikzpicture}[
    >=Stealth,
    node distance=0.65cm and 0.9cm,
    state/.style={
        draw=black,
        fill=white,
        rectangle,
        rounded corners=3pt,
        thick,
        font=\sffamily\scriptsize\bfseries,
        minimum width=2.4cm,
        minimum height=0.55cm,
        align=center
    },
    lbl/.style={
        font=\sffamily\tiny,
        fill=white,
        inner sep=1.5pt,
        align=center
    },
    edge/.style={
        draw=black!90,
        ->,
        thick,
        rounded corners=3pt
    }
]

    \node[state] (idle) {S\_IDLE};

    \node[state, below=0.9cm of idle] (wqe_post) {WQE\_POST};
    \node[state, below=0.6cm of wqe_post] (ring_sq) {RING\_SQ};
    \node[state, below=0.6cm of ring_sq] (wait_rpl) {WAIT\_RPL};
    \node[state, below=0.6cm of wait_rpl] (poll_rpl) {POLL\_RPL};
    \node[state, below=0.6cm of poll_rpl] (join) {JOIN};
    \node[state, below=0.9cm of join] (next) {NEXT};

    \node[state, left=1.0cm of wqe_post] (dat_rx) {DAT\_RX};
    \node[state, left=1.0cm of join] (dat_tx) {DAT\_TX};

    % --- EDGES ---
    \draw[edge] (idle) -- node[lbl, pos=0.45] {sw\_start} (wqe_post);
    \draw[edge] (idle.west) -| node[lbl, pos=0.25] {hw\_start} (dat_rx.north);

    \draw[edge] (dat_rx) -- node[lbl, above=1pt] {dma\_done} (wqe_post);

    \draw[edge] (wqe_post) -- node[lbl, pos=0.45] {post\_done} (ring_sq);
    \draw[edge] (ring_sq) -- node[lbl, pos=0.45] {ring\_done} (wait_rpl);
    \draw[edge] (wait_rpl) -- node[lbl, pos=0.45] {rcv\_completion} (poll_rpl);

    \draw[edge] (poll_rpl.east) -- ++(0.35,0) |- node[lbl, right, near start] {!seq\_match} ($(poll_rpl.east)+(0,0.18)$);
    \draw[edge] (poll_rpl) -- node[lbl, pos=0.4] {seq\_match or timeout} (join);

    \draw[edge] (wait_rpl.east) -- ++(0.85,0) |- node[lbl, left, pos=0.4] {!reply\_poll or\\timeout} (next.east);

    \draw[edge] (join) -- node[lbl, pos=0.45] {!demo\_act and \\prepost\_done} (next);
    \draw[edge] (join) -- node[lbl, below=6pt] {demo\_act \\and \\prepost\_done} (dat_tx);

    \draw[edge] (dat_tx.south) |- node[lbl, pos=0.75, above=1pt] {tx\_done} (next.west);

    \draw[edge] (next.east) -- ++(1.79,0) |- node[lbl, left=1pt, pos=0.3] {round\_done} (idle.east);

\end{tikzpicture}%
}
\caption{Consolidated FSM transition diagram for the \texttt{hh\_orchestrator} control unit.
In \texttt{WQE\_POST} the orchestrator constructs a 64-byte WQE and computes the local and remote
offsets inline, posting it to the SQ slot memory if the previous descriptor was not already
pre-posted. \texttt{RING\_SQ} rings the ERNIC's SQ producer-index doorbell and starts the RTT timer.
After waiting on an ERNIC completion in \texttt{WAIT\_RPL}, \texttt{POLL\_RPL} polls memory until a matching
sequence-number tag written by the remote endpoint indicates valid data; on match or timeout the
round-trip time is recorded. \texttt{DAT\_RX} and \texttt{DAT\_TX} carry the payload between PL and
the remote endpoint under \texttt{hw-loop}. The next round's WQE is posted as soon as the ERNIC
completion arrives, and execution synchronizes at \texttt{JOIN} to confirm that the background
pre-post has finished.}
\label{fig:hh_orchestrator_fsm}
\end{figure}

The application-layer logic is governed by the single-issue hardware-handshake orchestrator's state machine in Fig.~\ref{fig:hh_orchestrator_fsm}. The state machine optimizes round-trip latency through the pre-posting of work descriptors.

 The edge labels in Fig.~\ref{fig:hh_orchestrator_fsm} denote the condition or
completion event that triggers each transition; \texttt{!} denotes negation.
The two entry paths distinguish a software-initiated round (\texttt{sw-loop}) from a
hardware-stimulus round (\texttt{hw-loop}). Along the common path, completion signals advance the
FSM after WQE posting, doorbell delivery, and ERNIC transmission. The reply
sequence number determines whether the observed data belong to the current
round, while the timeout path prevents an unsuccessful transaction from
stalling the engine indefinitely.

The \texttt{JOIN} state synchronizes reply detection with the background
pre-post of the next WQE: \texttt{prepost\_done} indicates that the descriptor
is ready, while \texttt{demo\_act} selects whether the completed correction is
returned through \texttt{DAT\_TX}. Otherwise, the FSM proceeds directly to
\texttt{NEXT}, where the round status is committed. This overlap keeps
descriptor preparation outside the steady-state critical path and allows
reply handling to remain entirely in PL.

\begin{table}[!t]
\centering
\footnotesize
\caption{Overall resource utilization and sub-module breakdown on VPK120.}
\label{tab:fpga_resource}
\begin{talltblr}[
  entry = none,
  label = none,
  note{a} = {Overall utilization also includes 397 I/O pins (57\%), 2 DSP blocks in the ERNIC Core, and 1 MMCM (8\%).}
]{  colspec = {l r r r r r},
  colsep  = 4pt,
  row{1}  = {font=\bfseries\footnotesize, c},
  row{2-Z}= {font=\footnotesize},
}
\toprule
Module & LUT & LUTRAM & FF & BRAM & URAM \\
\midrule
\SetCell[c=6]{l, font=\bfseries\small} Overall Device Utilization \\
\midrule
Available              & 900,224 & 450,112 & 1,800,448 & 1,341 & 677 \\
Total Used\TblrNote{a} & 141,652 & 23,010  & 168,559   & 725   & 220 \\
Utilization (\%)       & 16\%    & 5\%     & 9\%       & 54\%  & 32\% \\
\midrule
\SetCell[c=6]{l, font=\bfseries\small} Component breakdown \\
\midrule
ERNIC Subsystem        & 73,077  & 4,254   & 52,207    & 172.5 & 156 \\
\quad ERNIC Core       & 67,748  & 4,254   & 44,679    & 165.5 & 20 \\
\quad HW Handshake     & 4,147   & 0       & 6,368     & 7     & 136 \\
\qquad Reply RAM       & 193     & 0       & 512       & 0     & 8  \\
\qquad RTT Trace       & 275     & 0       & 705       & 0     & 128 \\
Emb\_mem\_gen\_0       & 1,811   & 0       & 1         & 224   & 0  \\
Emb\_mem\_gen\_1       & 204     & 0       & 0         & 0     & 64 \\
Emb\_mem\_gen\_2       & 2,212   & 0       & 2         & 256   & 0  \\
\bottomrule
\end{talltblr}
\end{table}

The overall design consumes only 16\% of the device's LUTs and 9\% of its flip-flops, leaving room to insert additional logic modules on the same FPGA and to leverage its low-latency fabric interconnects. On-chip memory usage is intentionally larger to support cycle-accurate RTT profiling and demonstrations.
\begin{itemize}
    \item \textit{Send Queue (SQ) \& Queue-Pair Context \& Demo Syndromes:} The majority of Block RAM blocks (\texttt{Emb\_mem\_gen\_0/Emb\_mem\_gen\_1} and ERNIC Core) are mapped to hold the hardware SQ descriptor rings, and completion queues. This guarantees immediate access to RDMA descriptors without incurring off-chip access latencies.
    \item \textit{Reply Buffers:} Dedicated dual-ported UltraRAMs store incoming data, enabling memory polling at full memory bandwidth to minimize poll miss costs.
    \item \textit{RTT Trace \& Debug Space:} To support latency profiling, 128 URAM blocks in the HWHS module form an addressable debug trace space ($1,048,576$ entries $\times 16$ bits = 2~MiB). This allows continuous, cycle-accurate RTT measurement at line rate for up to one million consecutive RDMA rounds.
    \item \textit{Benchmarking syndromes:} Reference syndromes and expected replies are stored exclusively in \texttt{Emb\_mem\_gen\_2}, allowing the HWHS engine to access them efficiently for hardware-paced decoding benchmarks.
\end{itemize}

\subsubsection{Controller modes and cadences}

Two settings are varied independently in the benchmark: what machinery drives the sequence of rounds,
and how often a round is issued. Table~\ref{tab:appendix_loop_modes} lists the two controller
modes. \texttt{hw-handshake} moves the handshake into PL so that no software sits on the measured
request/reply path, while \texttt{hw-loop} additionally moves the round scheduling into PL, so the
measurement includes the time to move the payload between engines as it would be in a deployed
system.

\begin{table*}[!htbp]
\centering
\footnotesize
\caption{Controller modes used by the current demos and measurements.}
\label{tab:appendix_loop_modes}
\begin{tabular}{llll}
\toprule
Mode & Round driver & Posting and reply detection & Purpose \\
\midrule
\texttt{hw-handshake-sw-loop} (\texttt{sw-loop})
  & Host loop & HWHS in PL
  & Demo path and host-driven calibration \\
\texttt{hw-handshake-hw-loop} (\texttt{hw-loop})
  & PL pacer & HWHS in PL
  & Autonomous QPU-side and cadence experiments \\
\bottomrule
\end{tabular}
\end{table*}

Under \texttt{hw-loop} the PL pacer decides when the next syndrome is issued, and
Table~\ref{tab:pacer_types} gives the two cadences reported here. Both use the same Fibonacci-LFSR
interval generator, \(\textit{interval}=\textit{freq}+(\mathrm{LFSR}\mathbin{\&}\textit{span})\),
with tap sequence \([32, 22, 2, 1, 0]\) and a fixed seed for reproducibility.

\begin{table}[!htbp]
\centering
\footnotesize
\caption{Hardware-pacer cadences retained in the current evaluation.}
\label{tab:pacer_types}
\begin{tabular}{lll}
\toprule
Cadence & Interval & Interpretation \\
\midrule
\texttt{b2b} & No added wait & Continuous-service stress case \\
\texttt{random-delayed} & \qty{5}{\ms}--\qty{1347}{\ms} & Infrequent fallback \\
\bottomrule
\end{tabular}
\end{table}

For \texttt{b2b} (back-to-back), \textit{freq} and \textit{span} are zero, so the next syndrome is issued as soon as
the previous correction returns; this is the continuous-service stress case. For \texttt{random-delayed}, the
masked LFSR produces a random quiet period, modelling a nearby real-time decoder that handles the
common cases and calls out to the remote CPU or GPU decoder only for difficult tail events. The
programmed quiet period falls between rounds and is excluded from the reported RTT.

We verified the generator before use: replaying \(10{,}000\) intervals with the deployed parameters
(\qty{200}{\mega\hertz} clock, \(\textit{freq}=1{,}000{,}000\),
\(\textit{span}=\mathtt{0x0FFFFFFF}\), seed~1) gives a mean of \qty{677.847}{\milli\second} against a
uniform expectation of \qty{676.089}{\milli\second}, and the distribution across 20 equal-width bins
is flat to within sampling error.

\subsection{FPGA experiment results}\label{app:latency_analysis}

\begin{figure*}[!htbp]
\centering
\includegraphics[width=0.92\textwidth]{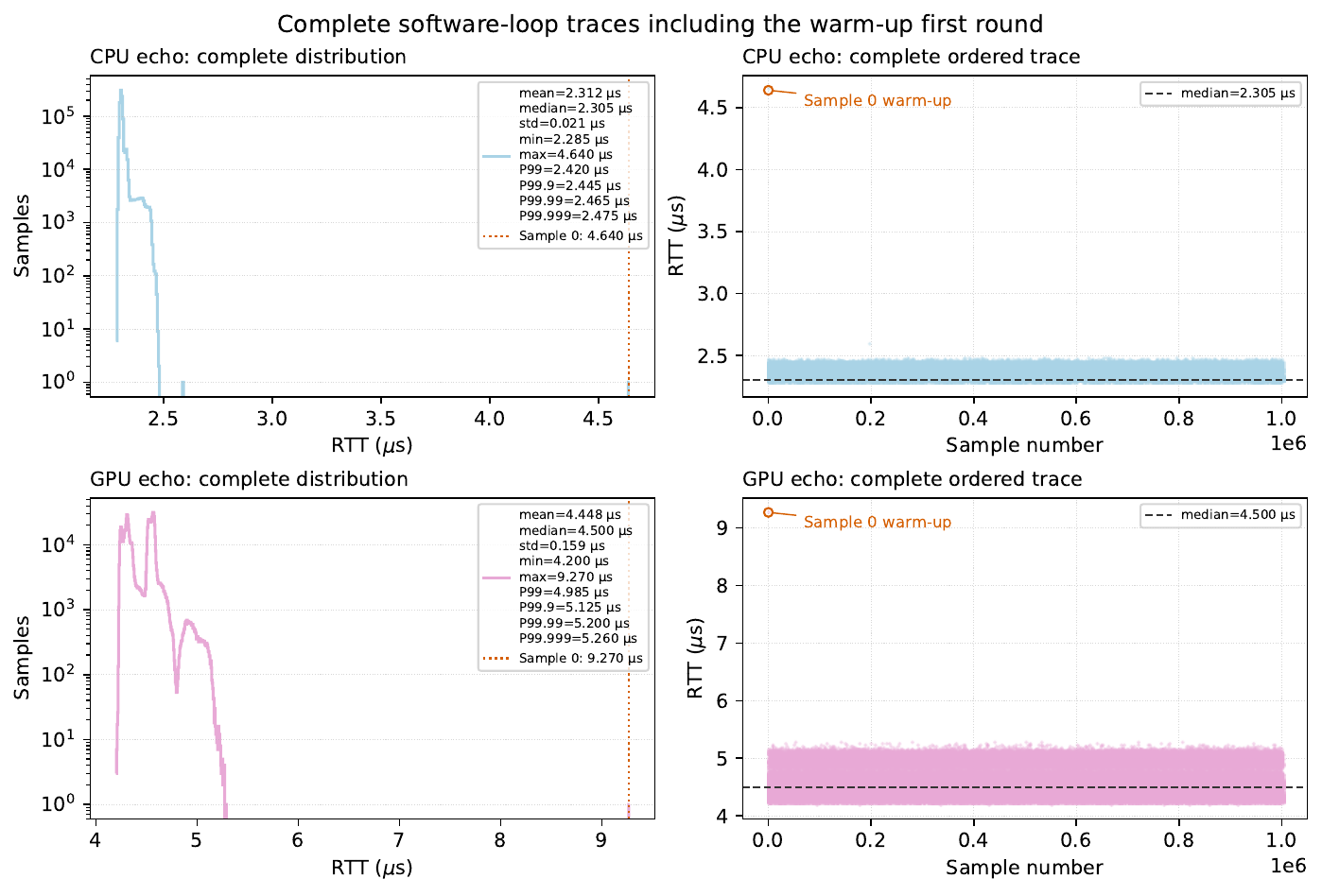}
\caption{Complete software-loop CPU/GPU echo traces. The left column shows the full distributions and the right column shows sample order. Round~0 is the one-time warm-up round containing the initial WQE and queue-pair settings.}
\label{fig:sw_loop_with_warmup}
\end{figure*}

The main text reports the primary \texttt{sw-loop} calibration in Fig.~\ref{fig:v_baseline} and Table~\ref{tab:v_responders}. As noted there, round~0 takes significantly longer than the steady-state mean, because it carries one-time endpoint initialization including the WQE and queue-pair settings: \qty{4.640}{\us} for CPU echo and \qty{9.270}{\us} for GPU echo, against steady-state medians of \qty{2.305}{\us} and \qty{4.500}{\us}. The complete ordered traces and distributions, including round~0, for the \texttt{sw-loop} experiments are shown in Fig.~\ref{fig:sw_loop_with_warmup}. Since this single-round initialization effect appears in every experiment, in the results reported subsequently (including Fig.~\ref{fig:hwhs_b2b_random-delayed_first} and Table~\ref{tab:hwhs_full_matrix}) we omit this round and show only the \(N-1\) steady-state rounds.

\subsubsection{Hardware-loop experiments}

Here we use \texttt{hw-loop} to understand the effect of the message-issuing cadence on round-trip latency; as mentioned, this mimics the sporadic decoding interval of a real setup. For \texttt{b2b} mode, each run executes \(N_{\mathrm{b2b}}=1{,}000{,}000\) rounds and for \texttt{random-delayed} mode, each run executes
\(N_{\mathrm{random-delayed}}=10{,}000\) rounds.\footnote{With a mean interval of \qty{676}{\milli\second} per \texttt{random-delayed} round, collecting 10{,}000 samples takes 1.88 hours excluding setup and result transfer; collecting 1{,}000{,}000 samples would take 188 hours.}

Table~\ref{tab:hwhs_full_matrix} reports the steady-state traces. CPU and GPU echo isolate the
transport, CPU and GPU Steane add the pre-compiled decoder, and qLDPC uses the generated GPU decoder. In \texttt{b2b} mode, the five responders have median RTTs of 2.470, 2.465, 4.575, 4.570, and
\qty{32.635}{\us}, respectively. CPU Steane adds almost nothing over CPU echo for this lookup
implementation, GPU Steane is \qty{0.005}{\us} faster than GPU echo, and the qLDPC kernel dominates its own
end-to-end RTT, consistent with the results reported in the main text.

\begin{table}[!htbp]
\centering
\scriptsize
\caption{Autonomous \texttt{hw-loop} steady-state benchmark matrix after removing Round~0. All
latency values are in microseconds.
These runs drive the loop from PL and therefore include the
PL-to-endpoint payload movement, so they are not directly comparable with the host-driven
\texttt{sw-loop} figures of Table~\ref{tab:v_responders} in the main text.}
\label{tab:hwhs_full_matrix}
\resizebox{\columnwidth}{!}{%
\begin{tabular}{llrrrrr}
\toprule
Responder & Cadence & Mean & Median & P99 & P99.9 & Max \\
\midrule
\texttt{cpu-echo} & \texttt{b2b}
  & 2.472 & 2.470 & 2.580 & 2.610 & 2.630 \\
\texttt{cpu-echo} & \texttt{random-delayed}
  & 2.494 & 2.490 & 2.630 & 2.740 & 2.820 \\
\addlinespace
\texttt{cpu-steane} & \texttt{b2b}
  & 2.471 & 2.465 & 2.570 & 2.605 & 2.635 \\
\texttt{cpu-steane} & \texttt{random-delayed}
  & 2.494 & 2.485 & 2.630 & 2.765 & 2.985 \\
\addlinespace
\texttt{gpu-echo} & \texttt{b2b}
  & 4.558 & 4.575 & 5.115 & 5.255 & 5.415 \\
\texttt{gpu-echo} & \texttt{random-delayed}
  & 4.543 & 4.485 & 5.115 & 5.320 & 5.505 \\
\addlinespace
\texttt{gpu-steane} & \texttt{b2b}
  & 4.611 & 4.570 & 5.160 & 5.295 & 5.445 \\
\texttt{gpu-steane} & \texttt{random-delayed}
  & 4.599 & 4.570 & 5.170 & 5.340 & 5.560 \\
\addlinespace
\texttt{gpu-qldpc} & \texttt{b2b}
  & 32.633 & 32.635 & 33.290 & 33.430 & 33.550 \\
\texttt{gpu-qldpc} & \texttt{random-delayed}
  & 32.561 & 32.570 & 33.215 & 33.465 & 33.815 \\
\bottomrule
\end{tabular}%
}
\end{table}

\subsubsection{Cadence effect on latency}
\label{app:b2b_random-delayed_equal}

\begin{figure*}[!htbp]
\centering
\includegraphics[width=0.98\textwidth]{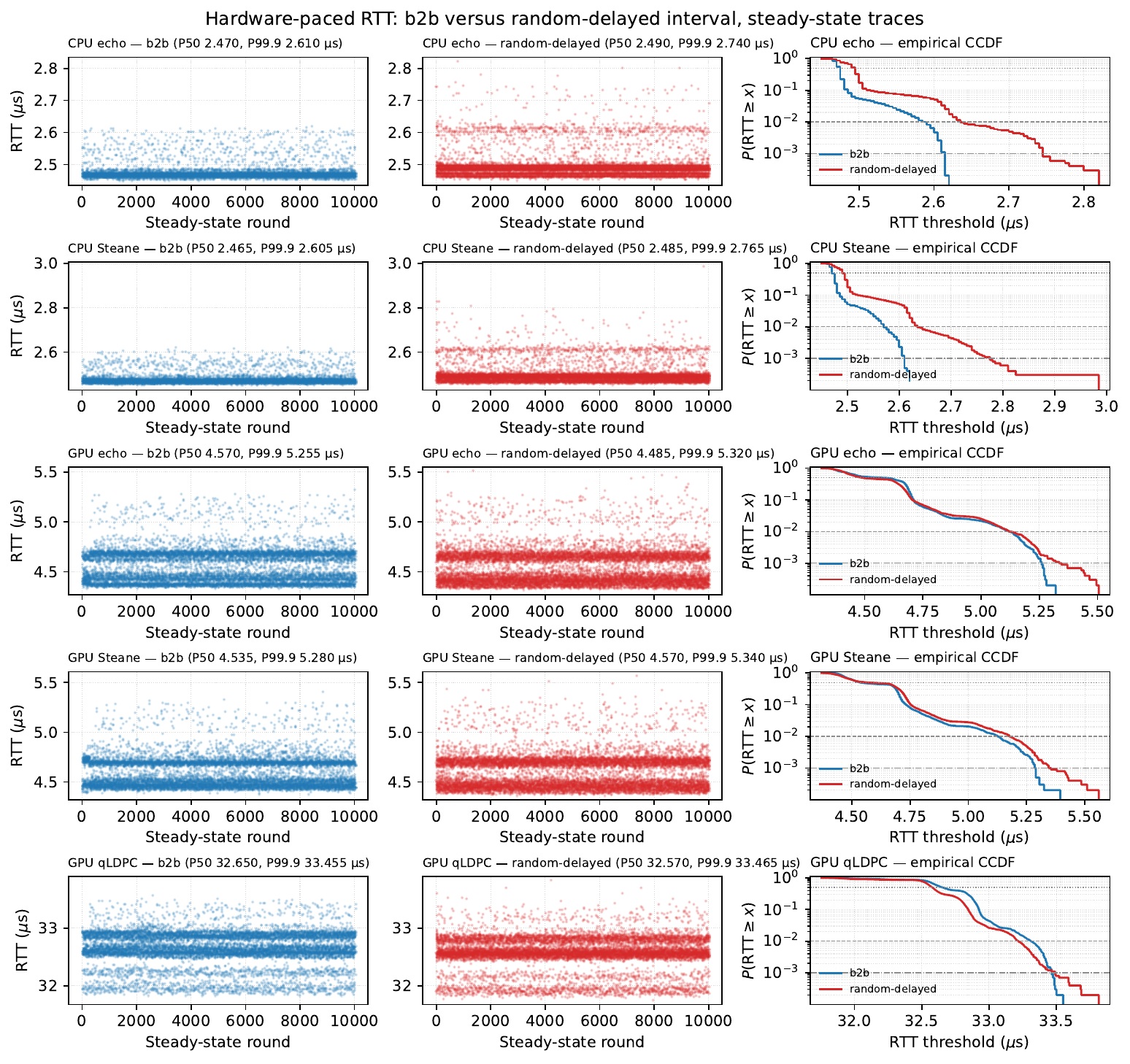}
\caption{Steady-state \texttt{b2b} (blue) and \texttt{random-delayed} (red) traces, one row per responder. Only the first \(N_{\mathrm{random-delayed}}-1\) rounds are plotted here. The P50 and P99.9 quoted in each panel include only the rounds plotted, so for \texttt{b2b} values come from the truncated subset and
can differ slightly from the full-run values in Table~\ref{tab:hwhs_full_matrix}. The first two columns plot round-trip time against round number and the third column overlays the two empirical CCDFs for that responder.}
\label{fig:hwhs_b2b_random-delayed_first}
\end{figure*}

Figure~\ref{fig:hwhs_b2b_random-delayed_first} shows the distributions for all five responders. The two
processor classes behave differently, and the difference is between the body of the distribution
and its tail.

On the CPU path, the whole distribution moves toward higher latency. Both responders are \qty{20}{\ns} slower in the median under \texttt{random-delayed}, while their tails widen further: P99.9 rises by \qty{130}{\ns} for echo and \qty{160}{\ns} for Steane. The identical median shift for echo and Steane places this cost on the long-gap path rather than on the decode itself.

On the GPU path, all three distributions move together slightly toward lower latency. Under \texttt{random-delayed}, the median decreases by \qty{90}{\ns}, \qty{0}{\ns}, and \qty{65}{\ns} for echo, Steane, and qLDPC, respectively. Their extreme tails nevertheless become heavier, with P99.9 increasing by \qty{65}{\ns}, \qty{45}{\ns}, and \qty{35}{\ns}. Thus, cadence shifts the body of the CPU and GPU distributions in different directions while widening the tail on both paths. These changes remain small relative to round trips of several microseconds.

Two hypothesized mechanisms are compatible with the measured shift:
\begin{itemize}
  \item \textbf{CPU C-state exit.} The \qty{676}{\milli\second} mean gap is long enough for an idle
  core to enter a deep state such as C6, which would be consistent with the effect being confined to the CPU
  responders. Against this, the responder is pinned and polls continuously, so deep-state residency
  may be limited, and the observed penalty is smaller than a typical full C6 exit.
  \item \textbf{P-state or frequency ramp.} Even with the polling core active, core or uncore
  frequency may fall during a quiet period. The time to return to the performance state would
  explain both the small central increase and the broader CPU tail without a deep C-state
  transition.
\end{itemize}

\end{document}